\documentclass[final,5p,times,twocolumn,authoryear]{elsarticle}

\usepackage{amssymb}

  \usepackage[caption=false,font=footnotesize]{subfig}

\usepackage{tabularx}
\usepackage{siunitx}
\usepackage{stmaryrd}
\usepackage{amsmath}
\usepackage{amssymb}
\usepackage{xcolor}

\usepackage{tabularx}
\usepackage[colorlinks,bookmarksopen,bookmarksnumbered,citecolor=red,urlcolor=red]{hyperref}
\usepackage{url}
\DeclareMathOperator{\logit}{logit}

\newcommand{\mr}{\mathrm}
\usepackage{amsfonts}
\usepackage{tabularx}
\usepackage{siunitx}

\journal{Journal of Systems and Software}

\begin{document}

\begin{frontmatter}



\title{Not all requirements prioritization criteria are equal at all times: \\A quantitative analysis}


\author[mymainaddress]{Richard Berntsson Svensson\corref{mycorrespondingauthor}}
\cortext[mycorrespondingauthor]{Corresponding author}
\ead{richard@cse.gu.se}

\author[mymainaddress,secondaddress]{Richard Torkar}
\ead{richard.torkar@gu.se}

\address[mymainaddress]{Department of Computer Science and Engineering, Chalmers and University of Gothenburg, Gothenburg, Sweden}
\address[secondaddress]{Stellenbosch Institute for Advanced Study, Stellenbosch, South Africa}

\begin{abstract}
Requirement prioritization is recognized as an important decision-making activity in requirements engineering   . Requirement prioritization is applied to determine which requirements should be implemented and released. In order to prioritize requirements, there are several approaches\slash techniques\slash tools that use different requirements prioritization criteria, which are often identified by gut feeling instead of an in-depth analysis of which criteria are most important to use. Therefore, in this study we investigate which requirements prioritization criteria are most important to use in industry when determining which requirements are implemented and released, and if the importance of the criteria change depending on how far a requirement has reached in the development process. We conducted a quantitative study where quantitative data was collected through a case study of one completed project from one software developing company by extracting 32,139 requirements prioritization decisions based on eight requirements prioritization criteria for 11,110 requirements. The results show that not all requirements prioritization criteria are equally important, and this change depending on how far a requirement has reached in the development process. For example, for requirements prioritization decisions before iteration\slash sprint planning, having high Business value had an impact on the decisions, but after iteration\slash sprint planning, having high Business value had no impact.
\end{abstract}



\begin{keyword}
Requirements prioritization \sep Bayesian analysis \sep empirical software engineering \sep construct validity




\end{keyword}

\end{frontmatter}


\section{Introduction}
Requirements Prioritization (RP) is an important decision making task in software development~\citep{Herrmann2008} where the objective is to determine, from a set of candidate requirements, which requirements are the most valuable and thus should be included in the product~\citep{Berander2005}, and in which order they should be implemented~\citep{Riegel2015}. Prioritizing requirements (i.e., determining the most valuable ones) involves making decisions based on one or several criteria, e.g., budget~\citep{Bukhsh2020}, time constraints~\citep{Bukhsh2020}, technical constraints (e.g., development cost and risk)~\citep{Riegel2015, SHAO2017, Pergher2013}, business aspects (e.g., market competition and regulations)~\citep{Pergher2013}, customer satisfaction~\citep{Pergher2013, SHAO2017}, or business value~\citep{Riegel2015, DANEVA2013}. The increasing number of requirements, both from internal (e.g., developers) and external (e.g., customers) sources, and from the availability of vast amount of data (big data) coming from digital networks connecting an increasing number of people, devices, services, and products~\citep{RBS2019}, makes RP even more difficult.

Several RP techniques have been introduced in the literature~\citep{Pergher2013, Achimugu2014, Hujainah2018, Bukhsh2020} to make RP accurate, efficient, and reliable~\citep{Bukhsh2020}. For example, RP techniques based on new technologies such as machine learning and repository mining~\citep{Pergher2013, Achimugu2016, SHAO2017} (following the trend of big data in requirements engineering), or RP techniques based on established RP concepts such as Analytical Hierarchy Process, Numerical Assignment, Planning Game, and Cumulative Voting~\citep{Bukhsh2020, rinkevicsT2013cv}.

Regardless if the RP techniques are based on new technologies or established concepts, all use one or several criteria when prioritizing requirements. However, all techniques have limitations, not only related to, e.g., scalability and requirements dependencies~\citep{Achimugu2014, SHAO2017}, but also due to assumptions about project context (e.g.,  the order in which requirements that should be prioritised are presented to the stakeholders, available information during RP, and how the RP process looks like)~\citep{Riegel2015, Bukhsh2020} and assumptions about which criteria should be used when prioritizing requirements, which is often decided based on gut feeling~\citep{Riegel2015}. Having a predefined set of criteria to be used in RP may lead to using misleading criteria~\citep{Riegel2015}, and thus making wrong\slash poor decisions. Hence, it is important to have flexible RP techniques where the used criteria are based on an in-depth analysis of which criteria are the most appropriate for a given context\slash project~\citep{Berander2005}. There are studies (e.g.,~\citep{Svensson2011, DANEVA2013, Jarzebowicz2020}) that have investigated which criteria are most commonly used in industry and\slash or most important\slash valuable when prioritizing requirements. Most (if not all) of these studies have investigated the RP criteria by asking industry practitioners for their subjective opinion concerning which criteria are most commonly used and\slash or most important when prioritizing requirements. However, the importance of various RP criteria and its content may be recalled differently among the practitioners due to memory bias~\citep{Zedtwitz2002} and details may quickly be forgotten~\citep{Baird1999}. Furthermore, reflection on purely experience-based memory recall (e.g., when asking practitioners about their subjective opinions) carries a high risk of drawing incorrect conclusions~\citep{Glass2002} and may result in emotional sessions rather than in constructive fact-based discussions. Therefore, to understand which RP criteria have an actual impact on RP decisions, an in-depth analysis~\citep{Berander2005} based on actual RP decisions are needed to avoid practitioners’ biases. To the best of our knowledge, no study has analyzed the actual outcome of RP decisions in industry to identify which RP criteria have an actual impact on RP decisions, or if the impact changes over time.

In this paper, we investigate which RP criteria have an actual impact in industry when determining which requirements are implemented and released to its customer and which ones are dropped. To this aim, we conducted a quantitative study of one completed project from one case company to investigate which criteria industry practitioners actually base their decisions on when prioritizing requirements, and if the criteria change depending on how far a requirement has reached in the development process. In order to investigate which RP criteria actually have an impact on RP decisions, we performed a quantitative study considering 32,139 RP decisions based on eight RP criteria for 11,110 features\footnote{The case company use the term feature for requirement} from one completed project with 14 software development teams and five cross-functional teams. That is, we collected quantitative data through a case study. The extracted data was analyzed by designing, comparing, validating, and diagnosing ordinal Bayesian regression models employing a \textsf{Sequential} likelihood. In addition, ordered categorical predictors were modeled as category-specific effects. Finally, to better understand how these effects vary over time a conditional effects analysis was conducted.

The results of this quantitative study show that not all RP criteria have an equal impact on RP decisions, and that the impact of a criterion changes depending on how far a requirement has reached in the development process. For example, having a high business value has an actual impact on RP early in the development process, high customer value has an impact in the middle, while being a critical requirement only has an impact at the end of the development process. Moreover, one out of eight used RP criteria, namely number of key customers who believed the requirement is important, had no impact on RP. Although the criterion dependency to other requirements had a significant impact on RP at one point in time, it did not matter if the requirement had dependencies to other requirements or not.

The remainder of this paper is organized as follows. Section~\ref{BGRW} presents related work, and an introduction to Bayesian Data Analysis. Section~\ref{RM} describes the design of our quantitative study, while Section~\ref{Res} presents the results. Section~\ref{Discussion} discuss the findings and Section~\ref{ValidityThreats} discloses the threats to the validity of our study. Finally, Section~\ref{Conclusion} gives a summary of the main conclusions.

\section{Background and related work}\label{BGRW}
This section presents related work on requirements prioritization. We conclude the section by providing a brief introduction to Bayesian data analysis.

\subsection{Requirements prioritization}
Several systematic literature reviews (SLRs) and systematic mapping studies have studied state-of-the-art in Requirements Prioritization (RP)~\citep{Herrmann2008, Kaur2013, Pergher2013, Achimugu2014, Hujainah2018, Thakurta2017, Bukhsh2020}. \citet{Herrmann2008} investigated RP techniques based on benefit and cost information and concluded that empirical validations of RP techniques where needed. \citet{Kaur2013} conducted an SLR and identified seven RP techniques that were compared and analyzed. The seven RP techniques were Analytic Hierarchy Process (AHP), value-oriented prioritization, cumulative voting, numerical assignment, binary search tree, planning game, and B-tree prioritization. The authors concluded that more work in the RP area is needed in order to improve the effectiveness---in terms of complexity and time consumption---of RP techniques. \citet{Pergher2013} performed a systematic mapping study focusing on empirical studies in RP. The authors identified that accuracy, time consumption, and ease of use were the most common criteria to use when evaluating RP techniques. Moreover, the results revealed that most studies in the RP area focus on RP techniques. \citet{Achimugu2014} conducted an SLR with the focus on RP techniques and their prioritization scales. The SLR identified 49 RP techniques that, in general, faced challenges related to time consumption, requirements dependencies, and scalability.

Later on, \citet{Hujainah2018} conducted an SLR to identify strengths and limitations of RP techniques. The results showed that RP is important for ensuring the quality of the developed system. In addition, 108 RP techniques were identified and analyzed based on, e.g., used RP criteria and limitations. In total, 84 RP criteria were used among the 108 RP techniques, where the criterion importance was the most frequently used. Moreover, the authors concluded that the existing RP techniques have limitations with regards to scalability, requirements dependencies, time consumption, and lack of quantification, which is in-line with the reported limitations in~\citep{Achimugu2014}. \citet{Thakurta2017} performed a systematic mapping study focusing on understanding RP artifacts, which included the objective of RP and factors that influence the overall RP process. In a recent SLR, \citet{Bukhsh2020} evaluated the existing empirical evidence in the RP area, which did not only include empirical evidence related to RP techniques. The results show that AHP is the most accurate and commonly used RP technique in industry. Most of the focus in the RP literature is on proposing, developing, and evaluating RP techniques, and comparing the performance of existing RP techniques~\citep{Pergher2013}. The most common approach to evaluate RP techniques is by empirically evaluate two or more RP techniques, where AHP is commonly used as one of them~\citep{Bukhsh2020}.

All RP techniques use one or several criteria for RP, where most of them use a fixed, predefined, set that are used during the RP process~\citep{Riegel2015}. However, the predefined criteria may not be suitable for all contexts. Thus, it is important to identify which criteria to use, and which ones are the most important to use given the context. \citet{Riegel2015} conducted an SLR to identify and categorize prioritization criteria. In total, about 280 prioritization criteria were extracted from the literature and categorized into six main categories: benefits, costs, risks, penalties and penalty avoidance, business context, and technical context and requirements characteristics. The most frequently mentioned RP criteria in the literature were: implementation effort, resource availability, implementation dependencies, business value, customer satisfaction, and development effort. \citet{Hujainah2018} identified 84 RP criteria where importance was the most used criterion among the identified RP techniques, followed by cost, business value, value, and dependency. \citet{Thakurta2017} identified several factors that influences RP, including requirements dependencies, software architecture, business value, and stakeholder roles.

Most of the identified RP criteria in the above literature comes from proposed RP techniques, and thus are selected based on gut feeling~\citep{Riegel2015} and not importance. There are studies, e.g.,~\citep{Svensson2011, DANEVA2013, Jarzebowicz2020}, that looked into which RP criteria are used\slash important in industry. \citet{Svensson2011} investigated how RP is conducted in industry and which criteria are used when prioritizing requirements. The results show that cost, value, customer input, and\slash or no criterion are the most commonly used criteria in industry for RP\@. In another study, \citet{DANEVA2013} found that the understanding of requirements dependencies is important for RP, and that the two most important RP criteria are business value and risk. \citet{Jarzebowicz2020} investigated agile RP in industry. The results show that business value is the most commonly used RP criterion, but other criteria such as complexity, stability, and interdependence are also used. However, these studies are based on the practitioners' subjective opinion about which RP criteria are important, and not on an in-depth analysis based on actual RP decisions.

The above indicates that the focus have been on comparing RP techniques and not on what we should measure, i.e., the criteria. Ultimately, in all analysis, what you measure and how you measure it, is more important than the actual analysis. To this end we focus on an analysis technique that allows us to take prior knowledge into account, handles disparate types of data, uses generative models, and quantifies uncertainty through probability theory, in order to investigate what effect different measurements have on RP.

\subsection{Bayesian data analysis}
Lately, many tools and probabilistic programming languages have been developed to tackle some of the challenges we face when designing more powerful statistical models. In our view, several things have improved. First, probabilistic programming languages, e.g., \texttt{Turing.jl} or \texttt{Stan}, have matured.\footnote{See \url{https://turing.ml} and \url{https://mc-stan.org}.} Second, resampling techniques based on Markov chain Monte Carlo (MCMC) have improved~\citep{brooks2011handbook}. Third, principled and transparent procedures for conducting Bayesian inference using the mentioned techniques now exist~\citep{talts2018arXiv,GabrySVBG17, gelman17likelihood,betancourt17hmc}, and are being continuously improved~\citep{vehtariGSCB21rhat}.

In our case we have decided to use the dynamic Hamiltonian Monte Carlo implementation that Stan provides~\citep{brooks2011handbook}. Mainly because it is considered to be the gold standard in the field, which many research groups use when benchmarking their algorithms. Additionally, compared to other MCMC implementations, Stan provides a plethora of diagnostics to ensure validity and reliability of the findings.

If we further contrast the above with how analyses are done in computer science and software engineering research today~\citep{gomesTFGFH19evo}, we feel that a principled Bayesian workflow serves us well. In short, issues such as the arbitrary $\alpha=0.05$ cut-off, the usage of null hypothesis significance testing and the reliance on confidence intervals have been criticized~\citep{ioannidis05false,MoreyHRLW2016CI,Nuzzo14errors,Woolston15pvalues}, and when analyzing the arguments, we have concluded that there is a need to avoid many of the issues plaguing other scientific fields.

In this paper we expect the reader to have knowledge regarding design of statistical models. In our particular case we will conduct linear regression, however our outcome (dependent variable) is of an ordered categorical nature (i.e., compared to a count the differences in value is not always equal), as are some of our predictors (independent variables)~\citep{burknerC20monotonic}.\footnote{Throughout the paper we use the terms variate, predicted variable, dependent variable, and outcome interchangeably. The same applies to the terms covariate, independent variable, and predictor. Since the paper's focus is on \textit{features} in requirements engineering, we refrain from using that term in connection to our statistical analysis.} To this end, we will design, compare, validate, and diagnose ordinal Bayesian regression models with the purpose of propagating uncertainty and making probabilistic statements by using a posterior probability distribution as explained by, e.g.,~\citep{burknerV19ordinal, furiaFT20bayes, torkarFFGGLE21pracsign, furiaTF2021applying}.

To summarize the development of statistical models in this paper we refer interested readers to the replication package.\footnote{\url{https://github.com/torkar/feature-selection-RBS} DOI: 10.5281\slash zenodo.4646845} However, we will provide a rudimentary overview of the steps involved (the below is a summary of what an interested reader can find in~\citep{furiaTF2021applying}).

First, model design begins with assumptions about the underlying data generation process (i.e., the likelihood). As the reader will see, the outcome in the data is ordered categorical, which leaves us with a number of options. We decide on the most appropriate likelihood by using ontological and epistemological arguments (see replication package). 

Second, once a rudimentary model has been developed one needs to set prior probability distributions (priors) on all parameters that we want estimated. To ensure that we are not overfitting, i.e., learning too much about the data, it is important that we check what the combinations of priors imply on our outcome. This is called prior predictive checks.

Third, once the model has been sampled, we check diagnostics to ensure that we have reached a stationary (stable) posterior probability distribution. If this is the case, we then check how well the model fits the empirical data; this is called posterior predictive checks.

Fourth, we next repeat the steps above and design more models. We then use Kullback-Leibler divergence~\citep{kullbackL51KLD} to decide on which model is the best (relative to other models) concerning out-of-sample predictions (i.e., which model deals best with new data); this is ultimately a cross-validation approach.

However, just because we have a `best' model, that does not mean that this will be true for all eternity. The approach above is iterative, and over time we learn more about the studied phenomenon and, hence, models will evolve over time when new evidence is added.

\section{Study design}\label{RM}
The aim of this study is to empirically evaluate the impact RP criteria have on decisions when prioritizing requirements, and if the criteria change depending on how far a requirement has reached in the development process. To address the aim of this study, a quantitative study where quantitative data was collected through a case study \citep{Wohlin2003} was used. Data was collected from one completed software project from one software developing company. The following research questions (RQ) provided the focus for the empirical investigation:

\begin{itemize}
  \item \textbf{RQ1:} Which of the used requirements prioritization criteria, by the case company, have an actual impact when determining which requirements should be implemented and released?
  \item \textbf{RQ2:} Does the impact of requirements prioritization criteria change depending on how far a requirement has reached in the development process? 
\end{itemize}

\subsection{Project selection criteria}\label{subsec:projectselection}
We conducted our analysis on one completed software project from one software developing company from our industry collaboration network. The software developing company has a large number of completed and ongoing projects. Thus, in order to select a project to be analyzed, four criteria were identified that needed to be satisfied:

\begin{itemize}
  \item \textit{Criterion 1: Completed project}. It was important for the studied project to be completed in order to analyze all requirements and decisions during the project's life cycle. Thus, we avoided projects with a short development time, e.g., projects that was only 50\% completed, since these projects would have an incomplete number of requirements and decisions made. 
  \item \textit{Criterion 2: More than one criterion}. About 280 different RP criteria have been identified in the literature~\citep{Riegel2015}, while other studies, e.g.~\citep{Svensson2011, Thakurta2017, Maalej2016}, have identified different criteria that are considered important in RP. Therefore, in order to analyze which criteria actually have an impact on the decisions in industry, it was important to analyze a project that used several different RP criteria.
   \item \textit{Criterion 3: Complete information}. We needed reliable data in order to produce a healthy dataset (the most important aspect in any statistical analysis is the data, not what approach one uses). To that end, all information and data about the requirements and the RP decisions needed to be documented and complete (i.e., no missing data\slash value\slash information about the requirements, RP criteria, or decisions made). This includes that all requirements' states should be documented, all used RP criteria including their values should be complete (i.e., no missing values), and all decisions (from RP) needed to be documented.
  \item \textit{Criterion 4: Large number of requirements and decisions}. In order to fully understand which RP criteria have an impact in industry, our studied project could not be a too simple example with only a few requirements and decisions made. Therefore, it was important that the studied project had a large number of requirements and decisions made, which could be seen as representative of a project at larger software company.
\end{itemize}

These four criteria allowed us to (i) identify a project that the company identified as a representative project of the case company, i.e., purposive sampling~\citep{ralphB20sampling} ensuring representativeness, (ii) discard projects having a short development time with few requirements and decisions, and (iii) discard projects with only one or a few RP criteria. A "gate-keeper" at the case company identified a suitable project that fulfilled all four criteria.

\subsection{Characteristics of the case company and the selected project}\label{subsec:casecompany}
The case company develops software for embedded products in a global consumer market. In the targeted organization of the case company (i.e., where the studied project belongs), there are about 1,000 employees in software development. The software development model used by the targeted organization is a continuous development model influenced by Scrum to allow for coordination with hardware and product projects. Requirement engineering is partly handled by the business department (as part of the cross-functional teams) and partly by the software development teams. The targeted organization use cross-functional teams that include customer representatives for key customers, which is either a representative from the real key customer or a customer proxy assigned by the business department. The cross-functional teams have full responsibility for defining, prioritizing, implementing, and testing features. A feature is developed and prioritized by one cross-functional team. In the targeted organization, 10--15 software development teams work in a typical project in several cross-functional teams for a duration of 2--3 years. A typical project has between 10,000 and 15,000 features.

\begin{figure*}
\includegraphics[width=\textwidth]{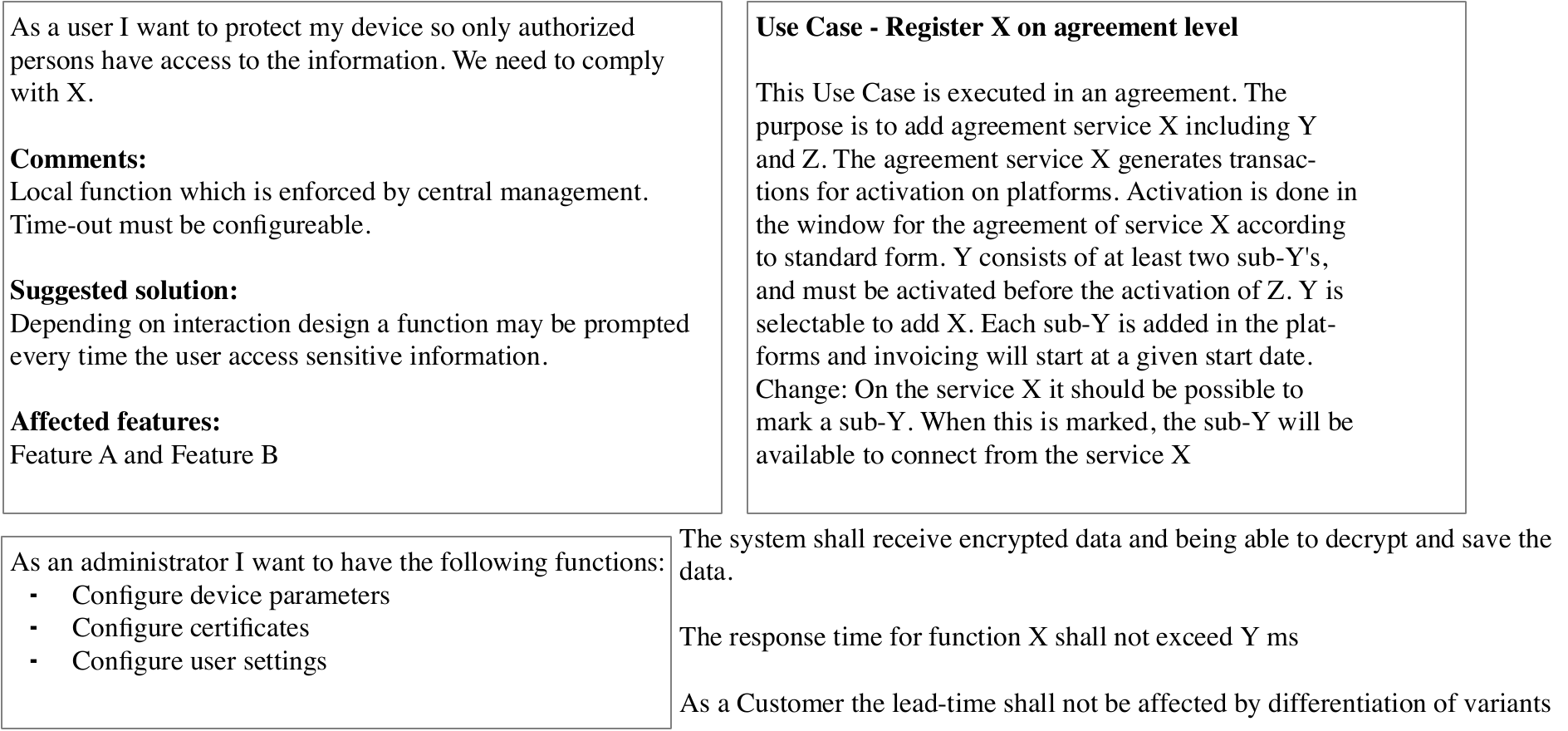}
\caption{Example of structure and what level of details the features had in the analyzed project}
\label{fig:examplereq}
\end{figure*}

The project in focus for this study, i.e., the studied project, is one of the targeted organization’s products. The studied project had a lead time close to three years from start to closure. In total, 14 software development teams in 5 cross-functional teams were involved in the development of the software for the embedded product. In total, the studied project had 11,110 features. Figure \ref{fig:examplereq} illustrates the structure of the features and what level of details the features had (written as user stories, natural language, and use cases). Note that the features in Fig.~\ref{fig:examplereq} are not the real features that were used in the analyzed project (due to confidentiality reasons, the used features are not allowed to be revealed). A feature could be in one of seven states, as shown in Fig.~\ref{fig:reqstate}.

The different states are described in Section~\ref{subsec:data}. The state of a feature shows how far the feature has reached in the development process. Before a feature was assigned to a state, a RP decision was made. All features were prioritized based on eight RP criteria by the responsible cross-functional team. The cross-functional team that developed and prioritized the feature was also responsible for collecting and recording the value for each RP criteria. In total, 32,139 RP decisions were based on the eight RP criteria. Table~\ref{tab:DatasetChar} provides a description of the studied software project. The eight RP criteria are described in Table~\ref{tab:FeatureVariables} together with the recorded values for each RP criterion.

\begin{table}
\caption{Characteristics of the analyzed software project}\label{tab:DatasetChar}
    \begin{tabularx}{\columnwidth}{XS}
\hline
\textbf{Characteristic} & {\textbf{\#}}\\
\hline
Features & 11,110\\
Decisions & 32,139\\
Requirements prioritization criteria & 8\\
Development teams & 14\\
Cross-functional teams & 5\\
\hline
\end{tabularx}
\end{table}

\subsection{Data extraction}\label{subsec:data}
\begin{figure*}
\includegraphics[width=\textwidth]{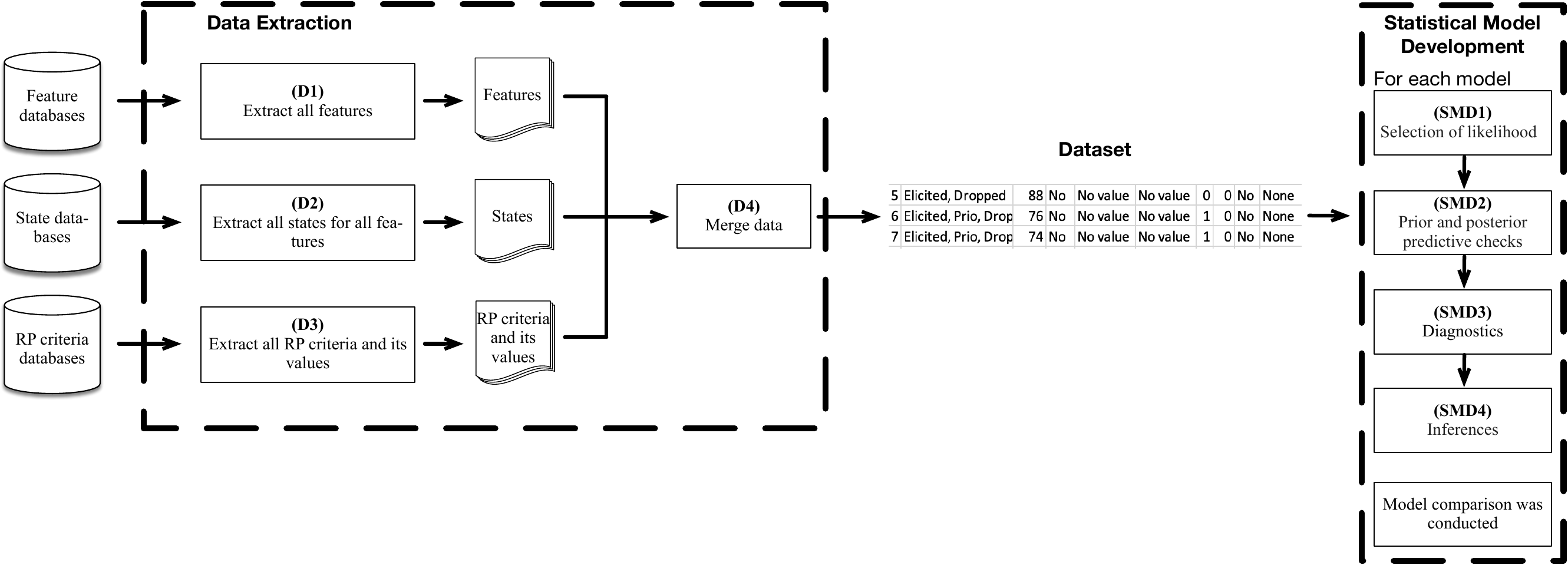}
\caption{Overview of the study design}
\label{fig:dataextracxtion}
\end{figure*}

We extracted data from three databases from the case company of the studied project. The first database contained all features of the studied project, the second database contained all states for all features, and the third database contained all RP criteria and its values for each feature of the studied project. Figure~\ref{fig:dataextracxtion} provides an overview of our data extraction steps, which are described below.

\textit{(D1) Extract all features}. The first step in the data collection and extraction phase was to extract all features that were ever considered from the completed project. For each feature, a unique ID (FeatureID) was extracted, which was used to link the feature to all RP decisions and state(s) the feature reached in the development process (see D2 below), and to all values it had for each requirement prioritization criteria (see D3 below). In total, 11,110 features were extracted.

\begin{figure*}
\includegraphics[width=\textwidth]{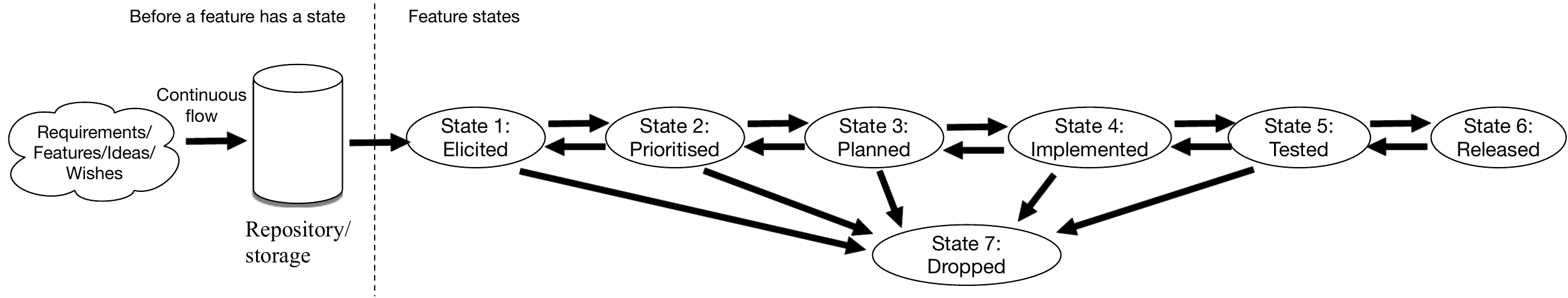}
\caption{Overview of feature states}
\label{fig:reqstate}
\end{figure*}

\textit{(D2) Extract all states for each feature}. When features are discovered it is not certain if the feature will be included in the product release. Available resources, scope, and lead-time limits the realization of any feature into the product. Therefore, to keep track of all features through the software development process, a feature can have one of seven states, namely: elicited, prioritized, planned, implemented, tested, released, or dropped (illustrated in Fig.~\ref{fig:reqstate}). The state of a feature shows how far the feature has reached in the development process. Before a feature is given a state (the first state is elicited), a decision (i.e., RP decision) was made to include that feature in the project. All extracted features from D1 reached at least the state of elicited, and thus is considered to be included in the project. Then, before a feature changes its state, a new RP decision based on eight criteria (see D3 below and Table~\ref{tab:FeatureVariables}) was made. A feature could move (backward or forward) from one state to another, meaning a feature could have been in one or several states. When extracting all states from the second database, the FeatureID was used to link each feature to all its states in the project. In this study, we only extracted the forward transitions of each feature. In total, 32,139 decisions were extracted. Figure~\ref{fig:reqstate} shows the different states for a feature, which are described below.

\textbf{State Elicited:} Each feature that has been through a pre-feasibility phase, and being prioritized (i.e., a decision is made to include the feature for the next step) reach the state Elicited.

\textbf{State Prioritized:} At regular intervals, features with the state of Elicited are being reviewed in a feasibility review for possible inclusion into the product. After the feasibility review, a decision (i.e., RP) is made. Features that are prioritized to be included get the state Prioritized.

\textbf{State Planned:} Features that have the state Prioritized are being reviewed in an analysis review. In the analysis review, each feature is analyzed based on, e.g., scope, adding details to the feature, estimations, and a more elaborate specification of the feature is created. After the analysis, the features are prioritized (i.e., a decision is made) to be included in the product or not. All features that are prioritized to be included in the product get the state Planned. These features are input for the design, coding, and iteration\slash sprint planning. Finally, these features are added to the product backlog.

\textbf{State Implemented:} From the product backlog, features are selected (i.e., prioritized) for development. When the features are developed, which includes technical design, coding, and unit tests, the features get the state Implemented.

\textbf{State Tested:} Although the implemented features include some testing, e.g., unit tests, a decision (RP) is made about which feature will be included for a more through testing process in order to ensure adequate level of quality before an implemented feature is released. Features that are selected for, and pass the testing, get the state Tested.

\textbf{State Released:} When all activities have been completed for the features with the state Tested, a decision (i.e., RP) is made about which features should be released. The features that are selected for being released get the state Released.

\textbf{State Dropped:} A feature can be rejected\slash dropped at any time in the process (until state Released). These features get the state Dropped. Dropped features are not deleted from the backlogs\slash repositories\slash storage to enable future analysis.

\textit{(D3) Extract all RP criteria and its values}. The FeatureID and States were used to extract the RP criteria and its values for each feature and its state(s). We extracted data from all RP criteria that were used when prioritizing a feature in this project. In total, eight different RP criteria were used each time a feature was prioritized. The extracted RP criteria are: team priority, critical Feature, customer value, business value, stakeholders, key customers, dependency, and architects' involvement, which are described in Table \ref{tab:FeatureVariables}. Note, we did not decide how many or which RP criteria were to be used in the analyzed project. This decision was made by the company before the project started.

\textit{(D4) Merge data}. After D3 was completed, we merged all extracted data from all three databases (D1--D3) using FeatureID and State. This allowed us to remove incomplete information, e.g., features without a state and empty values for the RP criteria.  Table~\ref{tab:FeatureVariables} provides an overview of all extracted data from D1--D3.

Let us now examine the outcome variable in particular, i.e., State. A \textit{State} is the state a feature has reached. Before a feature can reach the next higher state (e.g., moving from State Elicited to State Prioritized) or being dropped (i.e., moved to State Dropped), a RP decision is made. This RP decision is called a \textit{cutpoint}. Meaning, for our six states (Elicited, Prioritized, Planned, Implemented, Tested, and Released) there are five cutpoints, cutpoint $1$ is between states Elicited and Prioritized, cutpoint 2 between states Prioritized and Planned\ldots and cutpoint $5$ is between states Tested and Released. The result from the RP decision is either that a feature reached the next higher state or it is dropped. This result is called an \textit{outcome}. Meaning, in our statistical model, RP happens at five different cutpoints (decision points) that control the exits of states $1$--$5$, which decides whether a feature reaches the next higher state or being dropped. 

Due to non-disclosure agreements, the empirical data, e.g., FeatureID, variable names, and values, are not allowed to be revealed. Hence, we generated a synthetic dataset with describing names and values. The modifications of the real data include, changing the real FeatureID to a random ID without replacement from 1 to 11,110. Moreover, all variable names (column Variable in Table~\ref{tab:FeatureVariables}) were changed to descriptive and generic names that described the purpose of the variable, inline with current literature. In addition, the values (column Value(s) in Table~\ref{tab:FeatureVariables}) have been modified to descriptive values. For example, the values for the variable State are changed to names that describes the state of a feature. Section~\ref{subsec:descstat} provides descriptive statistics of the merged data.

\subsection{Descriptive statistics of the merged data}\label{subsec:descstat}

\begin{figure*}
\centering
\subfloat[]{%
  \includegraphics[scale=0.25]{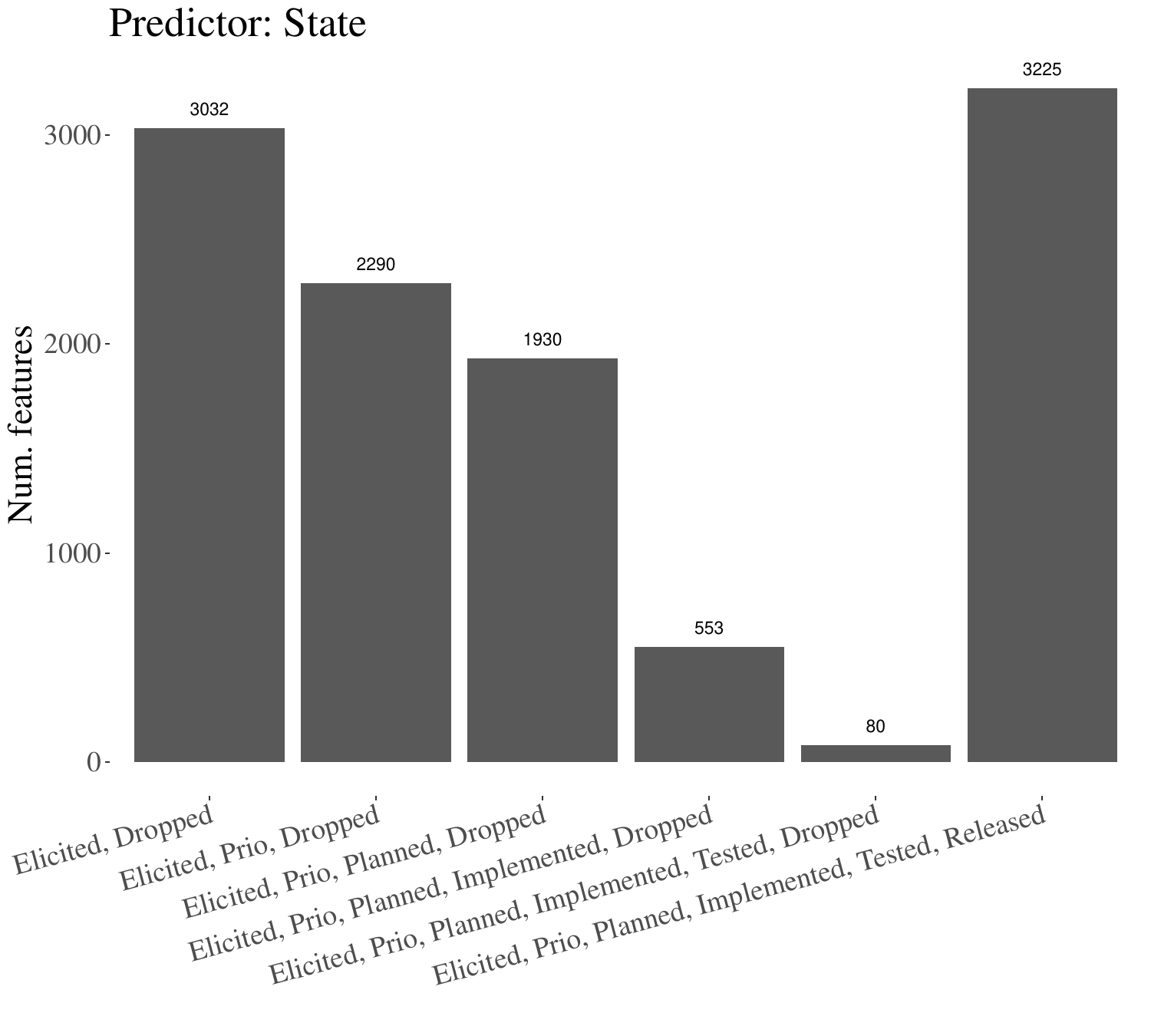}%
  \label{fig:hist-state}%
}\qquad
\subfloat[]{%
  \includegraphics[scale=0.25]{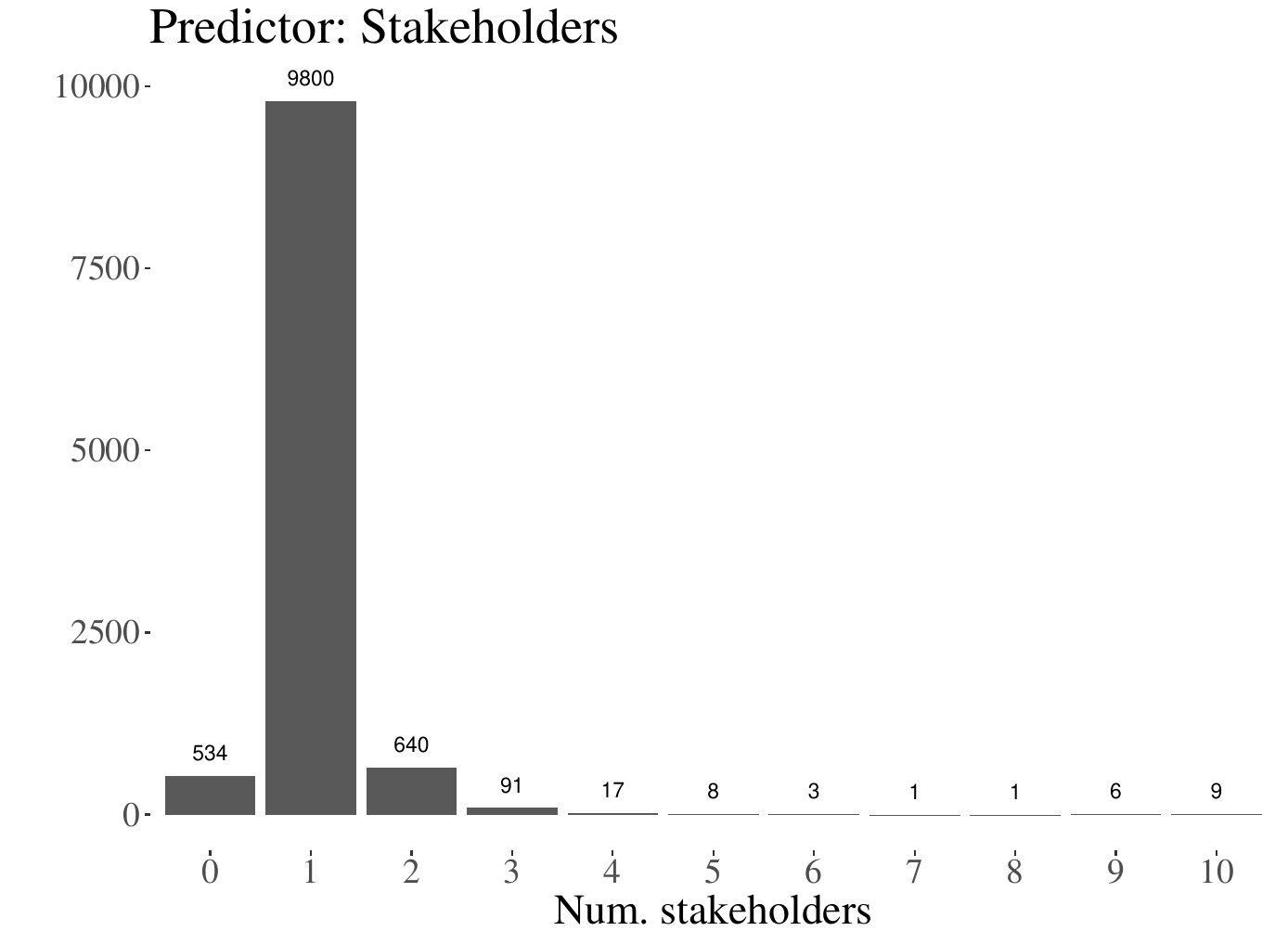}%
  \label{fig:hist-stake}%
}
\subfloat[]{%
  \includegraphics[scale=0.25]{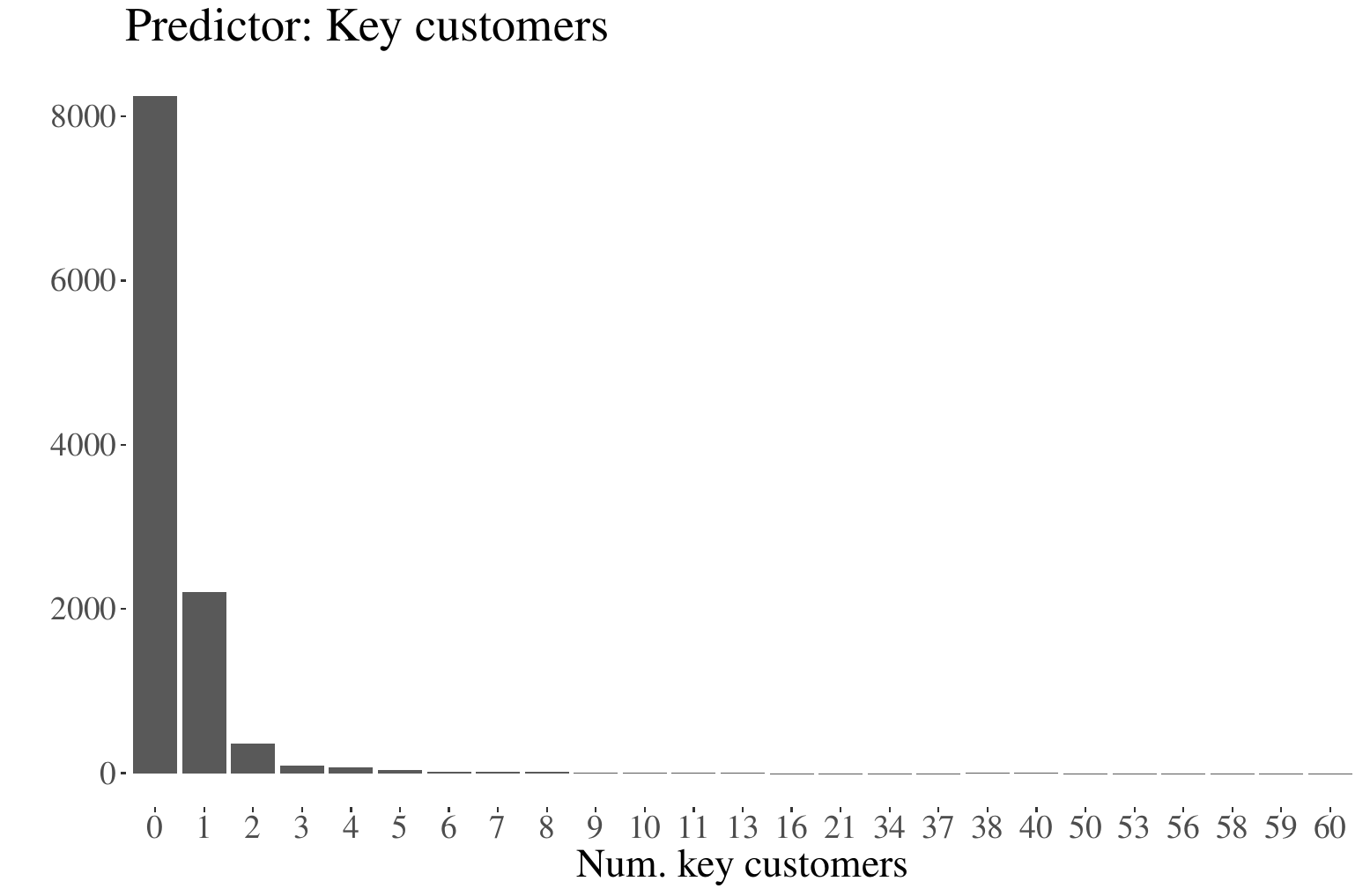}%
  \label{fig:hist-key-cust}%
}\qquad
\subfloat[]{%
  \includegraphics[scale=0.25]{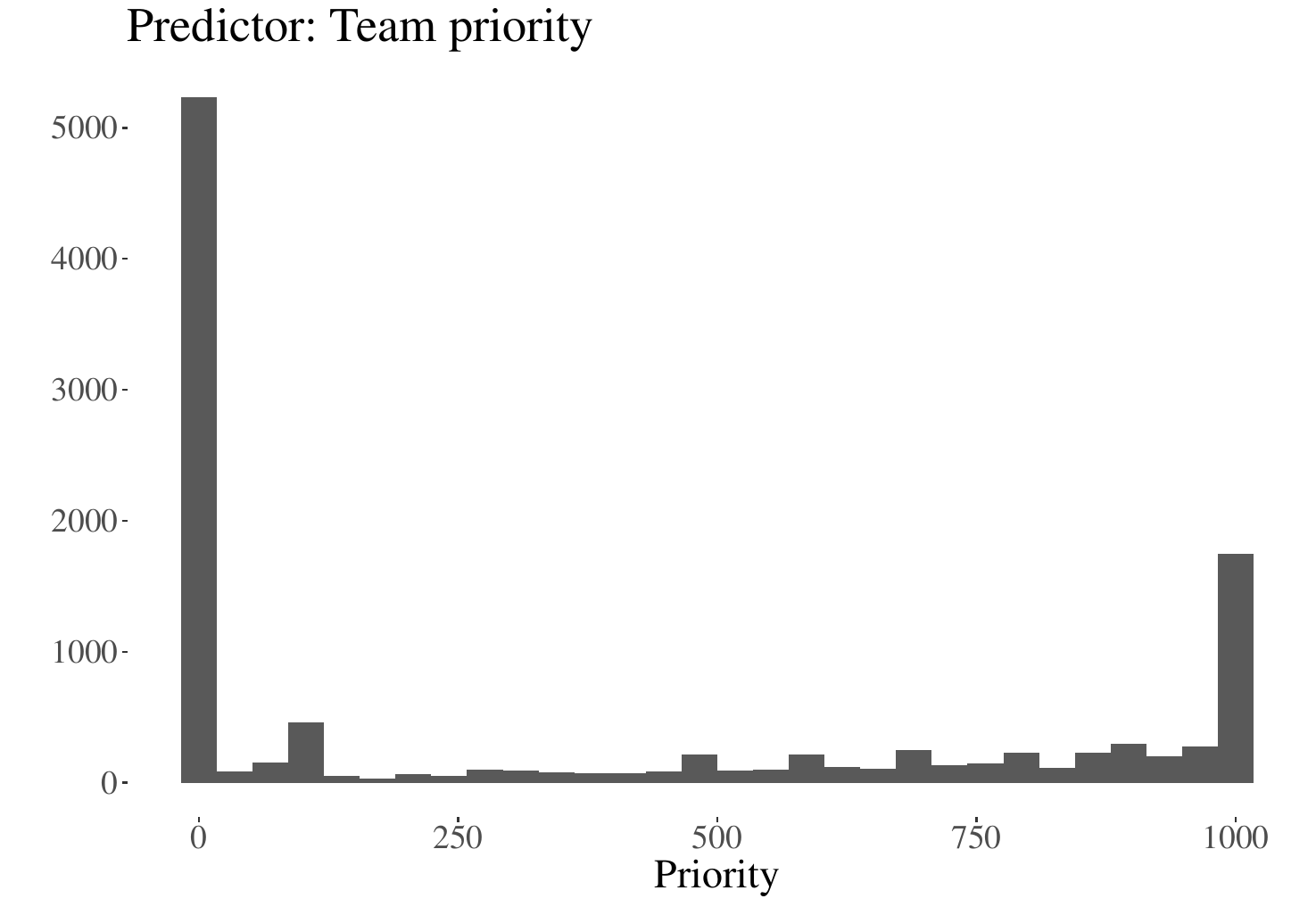}%
  \label{fig:hist-prio}%
}
\subfloat[]{%
  \includegraphics[scale=0.25]{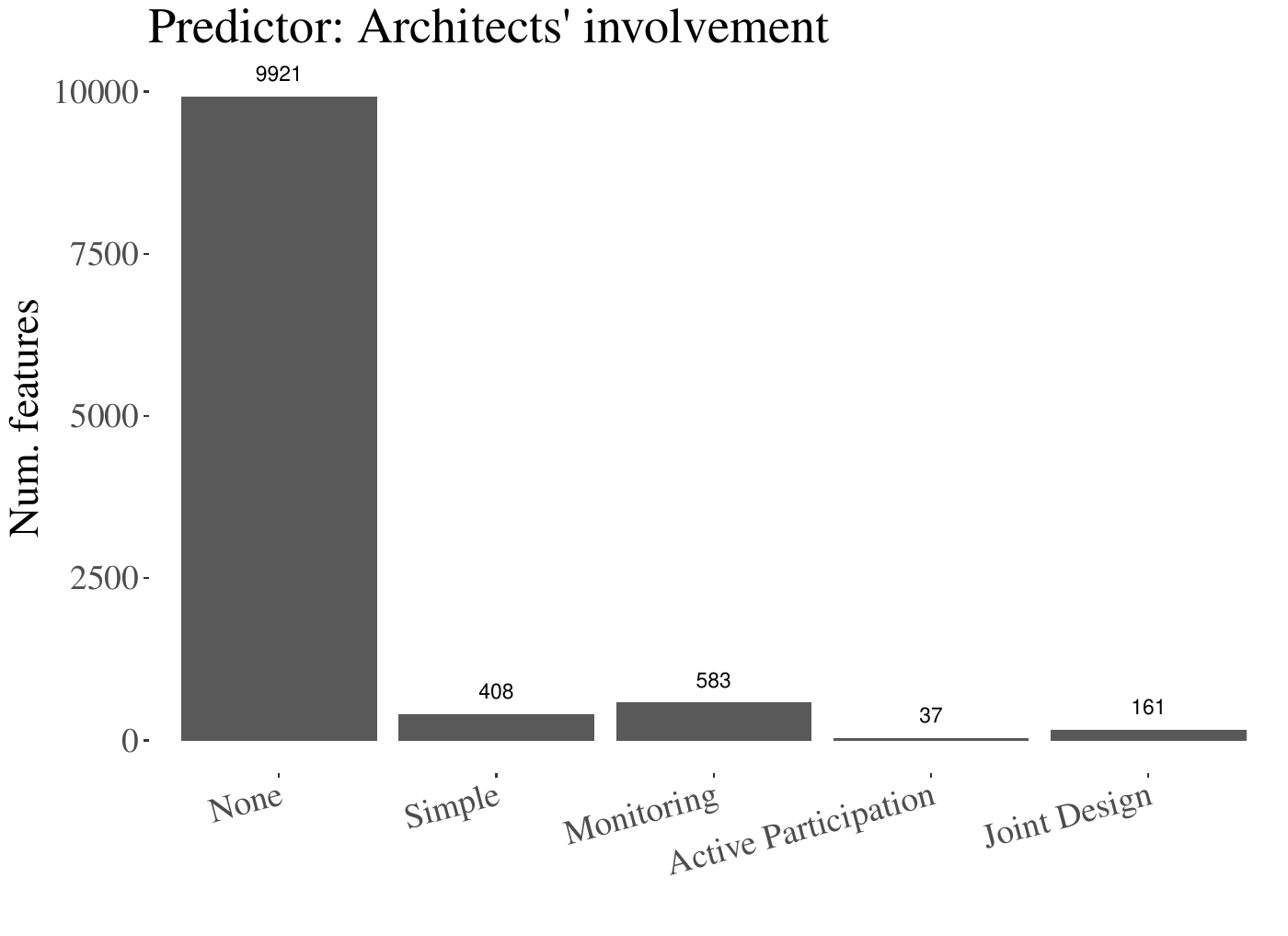}%
  \label{fig:hist-arch}%
}\qquad
\subfloat[]{%
  \includegraphics[scale=0.25]{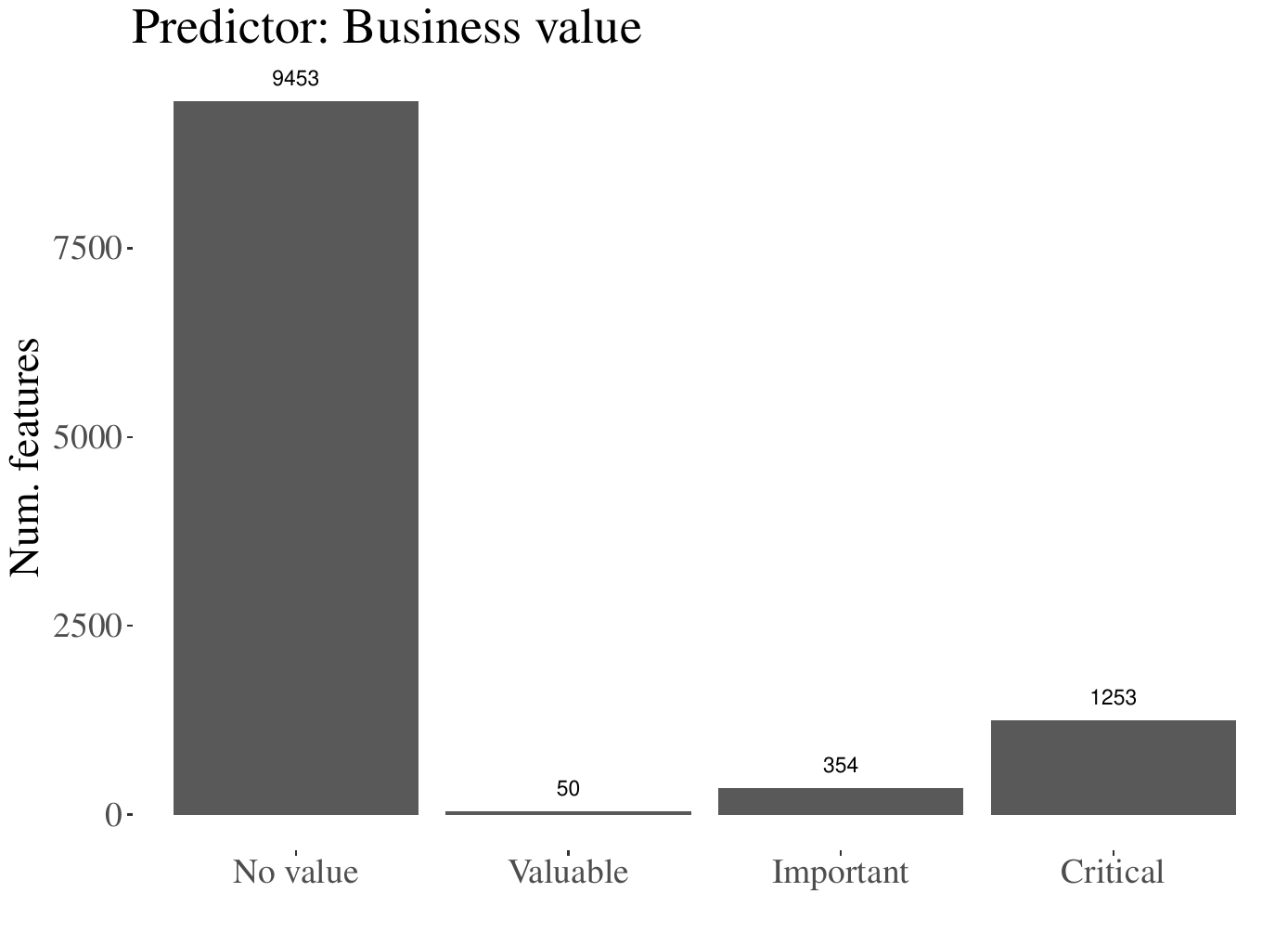}%
  \label{fig:hist-buss}%
}
\subfloat[]{%
  \includegraphics[scale=0.25]{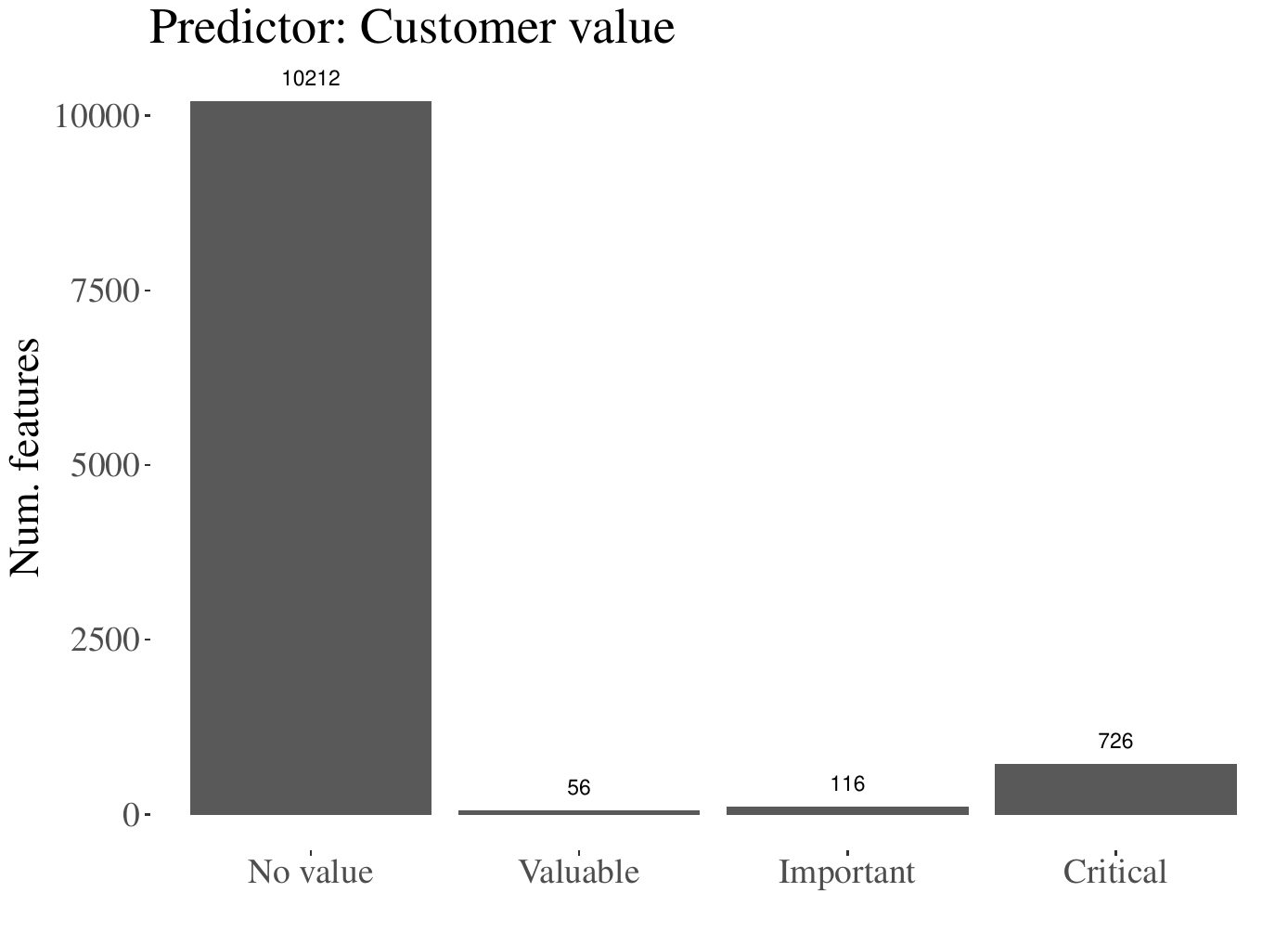}%
  \label{fig:hist-cust}%
}
\caption{Plots of outcome (a) and predictors (b--g), where e--g are ordinal independent variables.The $y$-axis represents the frequency. Two additional predictors are not plotted (\texttt{Dependency} and \texttt{Critical feature}).}\label{fig:p1}
\end{figure*}

Table~\ref{tab:FeatureVariables} provides a short description of the variables for each feature, while Fig.~\ref{fig:p1} presents their frequencies. For the variables \texttt{Dependency} and \texttt{Critical feature}, the answer was Yes\slash No and the ratio was 2004\slash 9106 and 1948\slash 9162, respectively.

\begin{table*}
\caption{Data dictionary. From left to right, variable name, encoding used in our model, possible values, description of the variable, and the type of variable. For type, $\mathbb{N}$ indicates a natural number, $\mathcal{O}$ indicates ordered data, and $\mathbb{Z}_2$ is binary data. The first row is our outcome variable, \texttt{State}.}\label{tab:FeatureVariables}
\begin{tabularx}{\textwidth}{lrXXl}
\hline
\textbf{Variable name} & \textbf{Code}&  \textbf{Possible values} & \textbf{Description} & \textbf{Type}\\
\hline
ID & n\slash a &1,\ldots,11110 & Unique ID for each feature & $\mathbb{N}^+$\\
\textbf{State} & state & Elicited (1), Prioritized (2), Planned (3), Implemented (4), Tested (5), Released (6), Dropped (7) & A feature's state that shows how far the feature reached in the process & $\mathcal{O}$\\
Team priority & prio & 0,\ldots,1000 & The relative priority a feature was assigned by the team & $\mathbb{N}^0$\\
Critical feature & crit & Yes\slash No & If a feature was considered to be critical for the product & $\mathbb{Z}_2$\\
Customer value & c\_val & No value, Valuable, Important, Critical & How valuable the feature was considered to be for customers & $\mathcal{O}$\\
Business value & b\_val & No value, Valuable, Important, Critical & The business value of the feature & $\mathcal{O}$\\
Stakeholders & sh & 0,\ldots,10 & Number of key internal stakeholders who considered a feature important & $\mathbb{N}^0$\\
Key customers & kc & 0,\ldots,60 & Number of key customers who considered a feature important & $\mathbb{N}^0$\\
Dependency &  dep & Yes\slash No & If a feature has a dependency to other features & $\mathbb{Z}_2$\\
Architects' involvement & arch & None, simple, monitoring, active participation, joint design & The needed level of involvement from architects in order to design\slash implement a feature & $\mathcal{O}$\\
\hline
\end{tabularx}
\end{table*}

If we examine our outcome \texttt{State} (Fig.~\ref{fig:hist-state}) we see that approximately 3,000 (out of 11,110) features are dropped already in the first state (Elicited $\shortrightarrow$ Dropped) and approximately 3,000 reach the final state (Elicited $\shortrightarrow \ldots \shortrightarrow$ Released). After the initial \texttt{State} $1$, fewer and fewer are dropped up to, and including, \texttt{State} $5$.

Concerning \texttt{Number of stakeholders} (Fig.~\ref{fig:hist-stake}), the absolute majority of the features have only one, while for \texttt{Number of key customers} (Fig.~\ref{fig:hist-key-cust}), most features have zero. Finally, concerning the variable \texttt{Priority} (Fig.~\ref{fig:hist-prio}), most features have zero in priority, and are then, more or less, spread out to priority 1000, where there is another peak.

Looking at the variables that eventually will be modeled as category-specific effects (i.e., they being ordered categorical) one can see that for \texttt{Architects' involvement} (Fig.~\ref{fig:hist-arch}) almost 90\% of the features do not have any architects involved. Additionally, for \texttt{Business value} and \texttt{Customer value} (Figs.~\ref{fig:hist-buss}--\ref{fig:hist-cust}) the distributions are comparable, where the first step, `No value', has been set 85--95\% of the time.

After gaining some insight concerning the variables, we next turn our attention to statistical model design where we design, compare, validate, and diagnose statistical models to conduct inferences. (Fig.~\ref{fig:dataextracxtion} provides an overview of our statistical model design.)

All steps in the analysis can be replicated by downloading the replication package, and preferably install \texttt{Docker}. The empirical data used in this manuscript is unfortunately not generally available due to an NDA\@. However, we have generated a synthetic dataset so anyone can follow the analysis step-by-step, and reach very similar results.

\subsection{Model design}
There are several ways to model ordered categorical (ordinal) data, but not until quite recently was it possible to use them easily in Bayesian data analysis. Software engineering, generally speaking, handles ordered categorical data by assuming that the conclusions do not depend on if a regression or ordinal model is used. The problem is, of course, that relying on an incorrect outcome distribution will lead to subpar predictive capabilities of the model~\citep{burknerV19ordinal}. This, in combination with the fact that effect size estimates will be biased when averaging multiple ordinal items, and that data can be non-normal, is something a researcher should want to handle~\citep{liddellK18ord}.

Today, we have at least three principled ways to model ordinal data: \textsf{Adjacent category}~\citep{burknerV19ordinal}, \textsf{Sequential}~\citep{tutz90seq}, and \textsf{Cumulative} models~\citep{walkerD67cm}. These models have been developed and refined in a Bayesian framework mostly because of needs from other disciplines, such as psychology~\citep{burknerV19ordinal}.

First, \textsf{Adjacent category} models can be used when predicting the number of correct answers to several questions in one category (think of a math module for the SAT or the PISA tests)~\citep{burknerV19ordinal}. We could perceive that our underlying data-generation process could be modeled this way.

\textsf{Sequential} models, on the other hand, assume that the outcome results from a sequential process and that higher responses are only possible if they pass lower responses; which is very much the case for our outcome \texttt{State}. 

Finally, \textsf{Cumulative} models assume that the outcome, e.g., observed Likert scale values, stems from a latent (not observable) continuous variable~\citep{burknerV19ordinal}.

In the case of \textsf{Sequential} models, we can model ordinal predictors as category-specific effects, while in \textsf{Cumulative} models, predictors are modeled as monotonic effects, the latter in order to avoid negative probabilities~\citep{burknerC20monotonic}. 

The main reason for modeling predictors as category-specific is to gain a more fine-grained view of the effect a predictor has on the outcome (i.e., how much does the predictor affect each outcome, \texttt{State} $1,\ldots,6$). In short, we want to model a predictor as an effect on $6$ ordered categories we use as \textit{outcomes}. The assumption that predictors have constant effects across all categories may be relaxed now, leading us to employ category-specific effects. 

As an example, consider our predictor \texttt{Architects' involvement}; it is quite likely that this predictor affects the outcome (\texttt{State} $1,\dots,6$) differently. Without using category-specific effects, this pattern would not be seen. In our case, we will later see that some category-specific effects are `significant'.

\textit{(SMD1) Selection of likelihood}. The first step concerning model design is often to decide which likelihood to use for inference, the \textsf{Cumulative}, \textsf{Sequential}, or \textsf{Adjacent-category}. This can be done by designing six statistical models and approximating their pointwise out-of-sample prediction accuracy (a measure of the out-of-sample fit). Doing this will allow us to receive estimates of how well each model handles new data (this, we would claim, is state of the art concerning model comparison, as introduced by~\citet{vehtariGG17loo}). In Table~\ref{tab:loo}, the result of the model comparison is presented, and it is clear that the \textsf{Sequential} model with predictors modeled as category-specific effects (where possible), has relatively speaking better out of sample prediction accuracy.

\begin{table}
    \centering
    \caption{From left to right, the model names, a short description indicating what type of model this is, and then the difference in expected log pointwise predictive density and the standard error. The abbreviation `cs' is short for category-specific effects, while `mo' is short for monotonic effects. The below is a ranked list so the top model is considered to have the best relative out of sample prediction capabilities. Comparing the first and second model, since $\Delta$elpd is $>4$x larger than the $\Delta$SE, one could claim that the models do not have similar predictive performance~\citep{pmlr-magnusson20a}. One can also see that using category-specific effects make a difference (1\textsuperscript{st} vs.\ 2\textsuperscript{nd} model) and that \textsf{Sequential} likelihood models are better than the alternatives (1\textsuperscript{st} and 2\textsuperscript{nd} vs.\ the other models).}
    \label{tab:loo}
    \begin{tabularx}{\columnwidth}{lXSS}
    \hline
         \textbf{Model} & \textbf{Description} & {$\Delta \textbf{elpd}$} & {$\Delta \textbf{SE}$}\\
         \hline
         $\mathcal{M}_{s[cs]}$ & \textsf{Sequential} w\slash cs & 0.0 & 0.0\\
         $\mathcal{M}_s$ & \textsf{Sequential} w\slash o cs & -43.6 & 10.8\\
         $\mathcal{M}_{ac}$ & \textsf{Adjacent-category} & -61.4 & 15.4\\
         $\mathcal{M}_{c[mo]}$ & \textsf{Cumulative} w\slash mo & -148.1 & 20.3\\
         $\mathcal{M}_c$ & \textsf{Cumulative} w\slash o mo & -148.6 & 20.2\\
         $\mathcal{M}_0$ & \textsf{Cumulative} w\slash o predictors & -3318.0 & 63.2\\
         \hline
    \end{tabularx}
    
\end{table}

If we examine Table~\ref{tab:loo}, there are several things it tells us. First, the model on the first row, $\mathcal{M}_{s[cs]}$, is different enough to warrant a first place. 
How do we know that? We can calculate the confidence interval (CI) between the first and second models, i.e., $-43.6 \pm 10.8 \cdot 2.576$ (2.576 is the z-score for the 99\% CI), which leads to $\mathrm{CI}_{99\%}[-74.42,-15.78]$. In summary, on the 99\%-level, the first model is significantly better than the second (since the difference does not cross zero). The only difference between the two first models is that we model predictors, when possible, as category-specific effects. 

Next, it is also notable that we do not see the same effect using a \textsf{Cumulative model} and modeling predictors as monotonic (rows 4--5). Finally, the last line is our null model ($\mathcal{M}_0$), which is a model that does not use any predictors and, thus, only models the mean. By looking at $\Delta$elpd, we see that adding predictors to our model (rows 1--5) has a clear effect compared to $\mathcal{M}_0$. Hence, the conclusion concerning the model comparison is that the \textsf{Sequential} model, using category-specific effects, is our target model for now $\mathcal{M} = \mathcal{M}_{s[cs]}$. Next, we need to set appropriate priors.

\textit{(SMD2) Prior and posterior predictive checks}. For our candidate model, we have several parameters in need of appropriate priors. One way to decide on priors is to make sure that the combination of all priors should be nearly uniform on the outcome scale and that impossible values should not be allowed.

Using a $\mathsf{Sequential}(\phi,\kappa)$ model we know that more probability mass could be set in the beginning (potentially all features could be dropped in \texttt{State} $1$), and then we should assign less probability mass for each following level in our outcome; we have six categories in our outcome, i.e., \texttt{State} $1$--$6$ (Fig.~\ref{fig:hist-state}).\footnote{Conventions for writing mathematical forms of \textsf{Sequential} models vary somewhat, but we will use $\mathsf{Sequential}(\phi,\kappa)$, where $\phi$ is our linear part and $\kappa$ the intercepts we want to estimate, i.e., the cutpoints between each step in the outcome. For the six levels in the outcome (\texttt{State} $1,\ldots,6$), we need $6-1 = 5$ cutpoints. The first cutpoint is the border between \texttt{State} $1$ and $2$, and the last cutpoint is the border between \texttt{State} $5$ and $6$. This way we can estimate the probability mass for each state.} The complete model design for $\mathcal{M}$ is thus,

\begin{eqnarray}
\mr{State}_i & \sim & \mathsf{Sequential}(\phi_i, \kappa) \\
\logit(\phi_i) & = & \beta_1 \cdot \mr{prio}_i + \beta_2 \cdot \mr{crit}_i + \beta_3 \cdot \mathrm{cs}(\mr{b\_val}_i) \\
& + & \beta_4 \cdot \mathrm{cs}(\mr{c\_val}_i) + \beta_5 \cdot \mr{sh}_i + \beta_6 \cdot \mr{kc}_i \\
& + & \beta_7 \cdot \mr{dep}_i + \beta_8 \cdot \mathrm{cs}(\mr{arch}_i) \\
\beta_1,\ldots,\beta_8 & \sim & \mr{Normal}(0, 1) \\
\kappa & \sim & \mr{Normal}(0, 2)
\end{eqnarray}

On the first line we assume that \texttt{State} is modeled using a \textsf{Sequential} likelihood. On Lines 2--4 we provide our linear model with all predictors and the parameters we want to estimate ($\beta_1,\ldots,\beta_8$). Ordered categorical predictors are modeled as category-specific effects, i.e., $\mr{cs}()$. As is evident, we model $\phi$ with a $\logit$ link function (Line 2), in order to translate back to the log-odds scale from the probability scale $(0,1)$.

Finally, on Lines 5--6, we set priors on our parameters. The intercept (cutpoints) priors for $\kappa$ are wider since we can expect them to vary more, while for our $\beta$ parameters $\mathcal{N}(0,1)$ might seem very tight, it still implies a prior variance of $\sigma^2 = (1 \cdot 8)^2 = 64$ for the model. 

\begin{figure}
    \centering
    \includegraphics[width=\columnwidth]{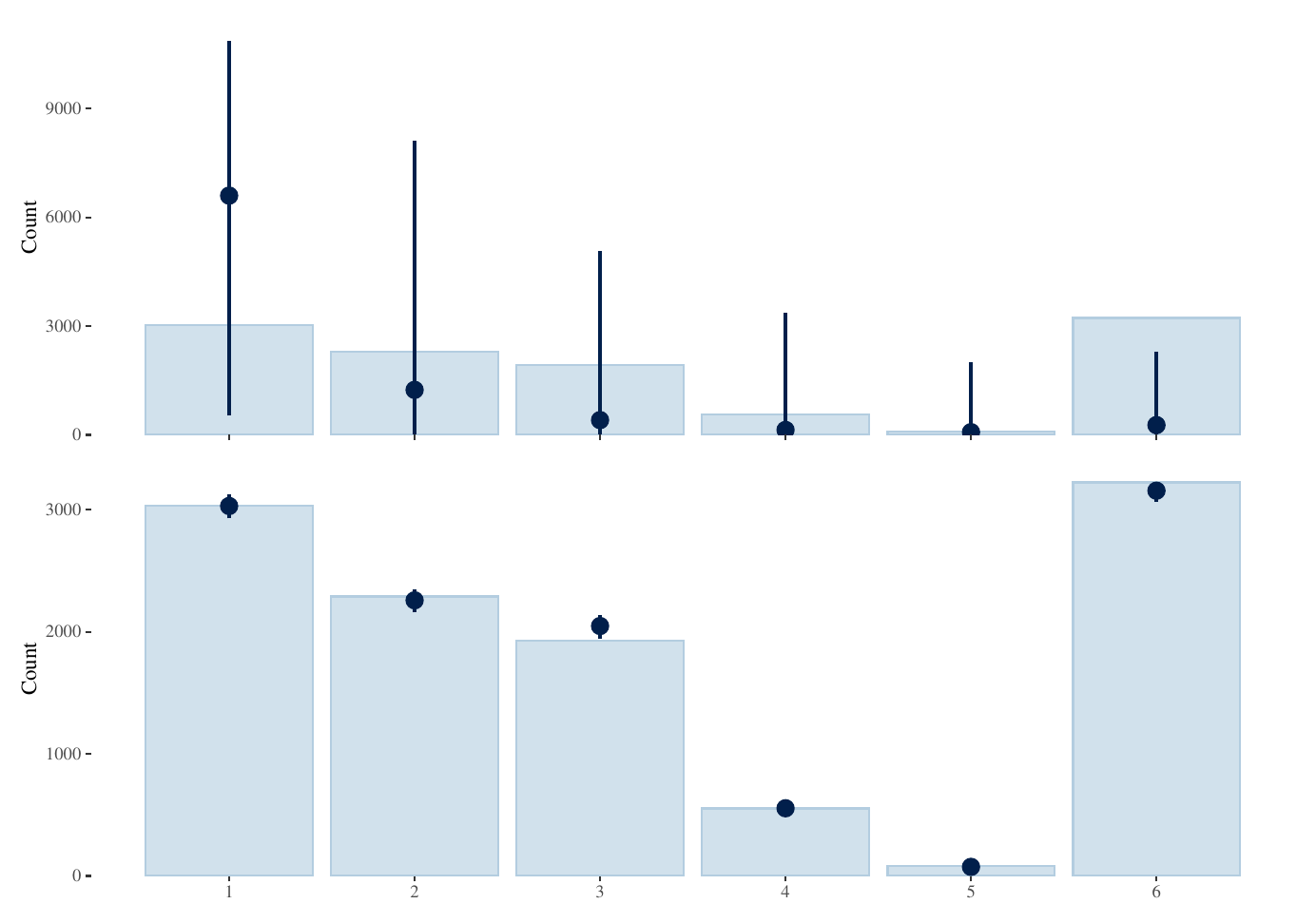}
    \caption{The top plot is the prior predictive check (we only sample from the priors and use no empirical data), while the bottom plot is the posterior predictive check (when we have used empirical data). The dots indicate the mean while the lines indicate the 95\% credible interval. The bars in both plots are our outcomes \texttt{State} $1,\ldots,6$ (note the different scales on the vertical axis). The combination of our priors (top plot) shows that we are expecting a negative slope giving more probability mass to the first category and then less for each following category. After making use of the empirical data (bottom plot) one can see a good fit (i.e., there is very little uncertainty around each mean and the means are placed close to the top of each bar). If we would see the same pattern as in the top plot and more uncertainty around each mean, that could imply the priors had too strong influence on the empirical data.}
    \label{fig:checks}
\end{figure}

A visual view can be given by sampling from the priors only, i.e., prior predictive checks, and with priors and data, i.e., posterior predictive checks (see Fig.~\ref{fig:checks} for a comparison).

\textit{(SMD3) Diagnostics}. When using dynamic Hamiltonian Monte Carlo we have a plethora of diagnostics, which we should utilize to ensure validity and efficiency of sampling~\citep{brooks2011handbook}. Validity concerns the degree one can trust the results, while efficiency is an indicator that, while we might be able to trust the results, the results could be imprecise.

Here follows a short summary of the most common diagnostics and the outcome of these diagnostics for $\mathcal{M}$.

There should be no divergences since it is an indication that the posterior is biased (non-stationary); it mainly arises when the posterior landscape is hard for HMC to explore (a validity concern). No divergences were reported.

Tree depth warnings are not a validity concern but rather an efficiency concern. Reaching the maximum tree depth indicates that the sampler is terminating prematurely to avoid long execution time~\citep{homanG14hmc}. No warnings were reported.

Having low energy values (E-BFMI) is an indication of a biased posterior (validity concern). No warnings were reported.

The $\widehat{R}$ convergence diagnostics indicates if the independent chains converged, i.e., explored the posterior in approximately the same way (validity concern)~\citep{vehtariGSCB21rhat}. It should converge to 1.0 as $n \rightarrow \infty$. The $\widehat{R}$ diagnostics was consistently $<1.01$, which is the current recommendation.

The effective sample size ($\mathrm{ESS}$) captures how many independent draws contain the same amount of information as the dependent sample obtained by the HMC algorithm, for each parameter (efficiency concern). The higher, the better. When $\mr{ESS} \approx 0.1$ one should start to worry, and in absolute numbers we should be in the hundreds for the Central Limit Theorem to hold. The $\mathrm{ESS}$ diagnostics was consistently $>0.2$.

Finally, Monte Carlo Standard Error (MCSE) was checked for all models. The MCSE is yet another diagnostic that reflects effective accuracy of a Markov chain by dividing the standard deviation of the chain with the square root of its effective sample size (validity concern).

In the replication package accompanying this paper the statistical validity and efficiency was checked for all models (see Sects.~2.*.3 in the replication package). All models passed all checks.

Having reached some confidence that the target model is representing the data generation process adequately, while assuring the validity and efficiency concerning the sampling, we next turn our attention to model validation (as opposed to validation of the output from the sampling algorithm).

First, for each model we conducted posterior predictive checks to see that the model captured the regular features of the data. Second, all model comparison was conducted using LOO~\citep{vehtariGG17loo}, which relies on approximate leave-one-out cross-validation. LOO was selected since it has good diagnostics warning the user of suspect results; compared to other techniques (i.e., WAIC, BIC, AIC), that lack such diagnostics.

\textit{(SMD4) Inferences}. The next section will provide results from 
the model by listing all parameter estimates (in our case the cutpoints, $\kappa$, are not relevant, but we shall focus on $\beta_1,\ldots,\beta_8$) and plot them. 

In particular, we will analyze the category-specific effects that were modeled. Does the fine-grained view, which the category-specific modeling of predictors provides us with, tells us a story about how each predictor, affects each outcome, i.e., \texttt{State} $1,\ldots,6$?

Finally, we will present a number of conditional effects. The latter concept is an excellent way to better understand the effect a specific predictor has on the six outcomes. Not only the size of the effect will be visible, but also how it varies depending on a number of factors.

The analysis follows the guidelines we present in previous work~\citep{furiaFT20bayes, furiaTF2021applying}, but for brevity, we do
not discuss here the application of the guidelines, but refer the interested readers to the replication package for details.

\section{Results}\label{Res}
Before we explain the concept of conditional effects, we will investigate the model's results as-is.

First, Table~\ref{tab:summary} consists of all parameters we are interested in (i.e., the $\beta$'s). All rows in bold indicate a significant effect, that is, the 95\% credible interval of an effect's distribution, does not cover zero (l-95\% and u-95\% columns in the table). 

What can we tell from Table~\ref{tab:summary}? First, \texttt{Priority}, \texttt{Criticality}, and \texttt{Dependency} have a positive effect (the higher the more likely to end up in a higher state), while the opposite is true for \texttt{Stakeholders}. Second, the predictor \texttt{Key customer} has very little predictive power. Third, for the predictors that were modelled as category-specific the picture is not that clear (i.e., they were modelled separately for each of the categories in the outcome \texttt{State}).

Looking at the first such predictor, i.e., \texttt{Business value} (\textbf{bus.~value[1,..,5]}), one can find three effects that are considered `significant': \textbf{bus.~value[1]}, \textbf{bus.~value[3]}, and \textbf{bus.~value[4]}. First, the higher the \texttt{Business value} in \texttt{State} $1$ and $3$, the likelier it is that it reaches those states, while the opposite holds for \texttt{State} $4$. In the latter case, a higher \texttt{Business value} leads, probabilistically speaking, more often to a requirement that will not reach \texttt{State} $4$. This  indicates  that  the  predictor  affects the outcome (our six states) differently.

\begin{figure*}
    \centering
    \includegraphics[scale=0.75]{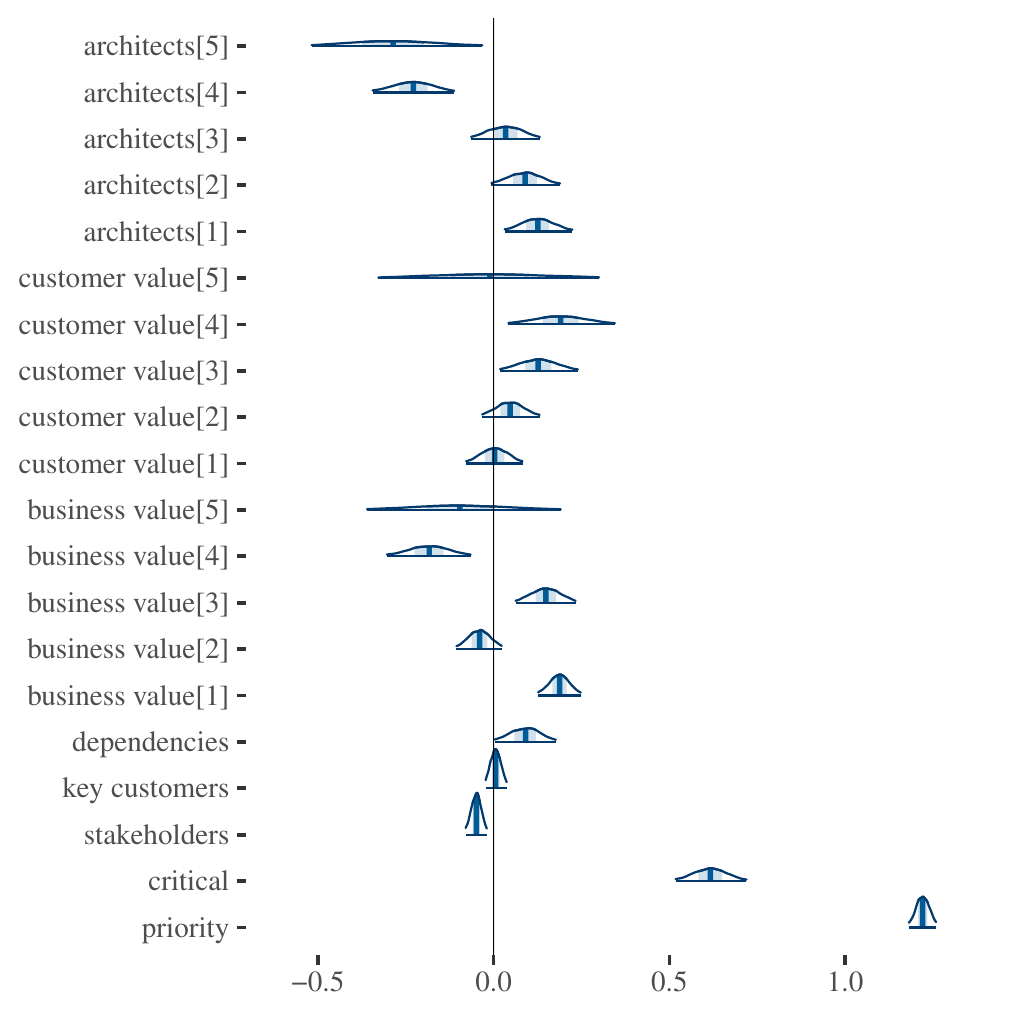}
    \caption{Density plot of all population-level (fixed effects). Examining the above plot, from bottom to top, we can claim that the first three parameters are significant and their 95\% CI do not cross zero (each density is cut off at 95\%). The fourth parameter, \texttt{Key customers}, is not significant. The fifth parameter, \texttt{Dependencies}, is significant $\mr{CI}_{95\%} = [0.01,0.18]$. For our three parameters modeled as category-specific: \texttt{Business value}, \texttt{Customer value}, and \texttt{Architects' involvement}, some categories are visibly not crossing zero. The rows \texttt{bus.~value[1,\ldots,5]}, \texttt{cust.~value[1,\ldots,5]}, and \texttt{arch.~involv.[1,\ldots,5]}, are the five cutpoints which we use to estimate the deviation on each outcome (\texttt{State} $1,\ldots,6$). Since we have six outcomes we have $6 - 1 = 5$ cutpoints (i.e., the borders between the six outcomes, \texttt{State} $1,\ldots,6$).}
    \label{fig:mcmc_areas}
\end{figure*}

Figure~\ref{fig:mcmc_areas} provides a visualization of the table that is, perhaps, more straightforward to understand. In Fig.~\ref{fig:mcmc_areas}, each parameter's posterior probability distribution, with 95\% credible intervals, is plotted. This, perhaps, allows the reader to gain more insight, compared to only looking at Table~\ref{tab:summary}.

\begin{table*}
    \centering
    \caption{Summary of population-level (fixed) effects. From left to right, the name of the effect, estimate, estimation error (a parameter's posterior standard deviation), and lower and upper 95\% credible intervals. Significant effects (95\%) are in bold and left-aligned. The rows \textbf{bus.~value[1,\ldots,5]}, \textbf{cust.~value[1,\ldots,5]}, and \textbf{arch.~involv.[1,\ldots,5]}, are the five cutpoints which we use to estimate the deviation on each outcome (\texttt{State} $1,\ldots,6$). Since we have six outcomes we have $6 - 1 = 5$ cutpoints (i.e., the borders between the six outcomes, \texttt{State} $1,\ldots,6$).}
    \label{tab:summary}
    \begin{tabularx}{\textwidth}{XSSSS}
    \hline
    \textbf{Effect} & \textbf{Estimate} & \textbf{Est.\ Error} & \textbf{l-95\% CI} & \textbf{u-95\% CI} \\
    \hline
\textbf{priority}     &   \textbf{1.22}    &  \textbf{0.02} &    \textbf{1.18} &    \textbf{1.26} \\
\textbf{criticality}        &   \textbf{0.62}    &  \textbf{0.05} &    \textbf{0.52} &    \textbf{0.71} \\
\textbf{stakeholders}       &   \textbf{-0.05}   &  \textbf{0.02} &   \textbf{-0.08} &   \textbf{-0.02} \\
key customers       &   0.01    &  0.02 &   -0.02 &    0.04 \\
\textbf{dependency}         &   \textbf{0.09}    &  \textbf{0.04} &    \textbf{0.01} &    \textbf{0.18} \\
\textbf{bus.~value[1]}    &   \textbf{0.19}    &  \textbf{0.03} &    \textbf{0.13} &    \textbf{0.25} \\
bus.~value[2]    &  -0.04    &  0.03 &   -0.10 &    0.02 \\
\textbf{bus.~value[3]}    &   \textbf{0.15}    &  \textbf{0.04} &    \textbf{0.06} &    \textbf{0.23} \\
\textbf{bus.~value[4]}    &  \textbf{-0.18}    &  \textbf{0.06} & \textbf{-0.31} & \textbf{-0.06} \\
bus.~value[5]    &  -0.09    &  0.14 &   -0.35 &    0.19 \\
cust.~value[1]    &   0.00    &  0.04 &   -0.07 &    0.08 \\
cust.~value[2]    &   0.05    &  0.04 &   -0.04 &    0.13 \\
\textbf{cust.~value[3]} &   \textbf{0.13}    &  \textbf{0.06} &    \textbf{0.02} &    \textbf{0.24} \\
\textbf{cust.~value[4]} &   \textbf{0.19}    &  \textbf{0.08} & \textbf{0.04} & \textbf{0.35} \\
cust.~value[5]    &  -0.01    &  0.16 &   -0.33 &    0.32 \\
\textbf{arch.~involv.[1]} & \textbf{0.13} & \textbf{0.05} & \textbf{0.03} & \textbf{0.22} \\
\textbf{arch.~involv.[2]} & \textbf{0.09} & \textbf{0.05} & \textbf{0.00} & \textbf{0.18} \\
arch.~involv.[3] &   0.03    &  0.05 &   -0.06 &    0.13 \\
arch.~involv.[4] &  -0.23    &  0.06 &   -0.34 &    0.11 \\
\textbf{arch.~involv.[5]} &  \textbf{-0.29} & \textbf{0.13} & \textbf{-0.52} & \textbf{-0.03} \\
    \hline
    \end{tabularx}
\end{table*}

Next, even though we are not very interested in an effect's point estimate \textit{per se}, let us take \texttt{Stakeholders}, as an example, i.e., $\mu = -0.05$ $\mr{CI}_{95\%}[-0.08,-0.02]$. Recall from Sect.~\ref{subsec:descstat} that \texttt{Stakeholders} could vary $(0,\ldots,10)$ and was used to indicate how many stakeholders a particular feature had. First, we transform the value using inverse $\logit$, since the model used a $\logit()$ link function, i.e., $\exp(-0.05)/(\exp(-0.05) + 1) = 0.49$.

In order to improve sampling, all variables, where appropriate, were centered and scaled, i.e., for all values, we removed the variable's mean ($\mu_x=1.05$), and then scaled each value by dividing it with the variable's standard deviation ($\sigma_x = 0.53$). Hence, to receive the original scale we do the opposite, i.e., $0.49 \cdot 0.53 + 1.05 = 1.31,{}\mr{CI}_{95\%}[1.306,1.314]$. In short, the model estimates that, on average, we have $1.31$ stakeholders per feature, with a 95\% credible interval of $[1.306,1.314]$.

If we next take customer value as an example, we can claim that customer value has a positive effect on the third and fourth cutpoints, i.e., the cutpoints between \texttt{States} $3$\slash $4$ and between $4$\slash $5$, are pushed up, leading to more probability mass being assigned to the lower levels, i.e., \texttt{State} $\leq4$ (\texttt{Elicited, Prio, Planned, Implemented, Dropped}, and below). This detail would have been impossible to notice without modeling the effect as category specific.\footnote{In the replication package one can see that not modeling category-specific effects would miss that architects' involvement actually has some significant effects.} However, what is even more interesting is the fact that we have a joint posterior probability distribution for all effects, i.e., it is possible to see how each effect varies when fixing all other covariates to their mean or reference category.

\subsection{Conditional effects}
\begin{figure}
    \centering
    \includegraphics[width=\columnwidth]{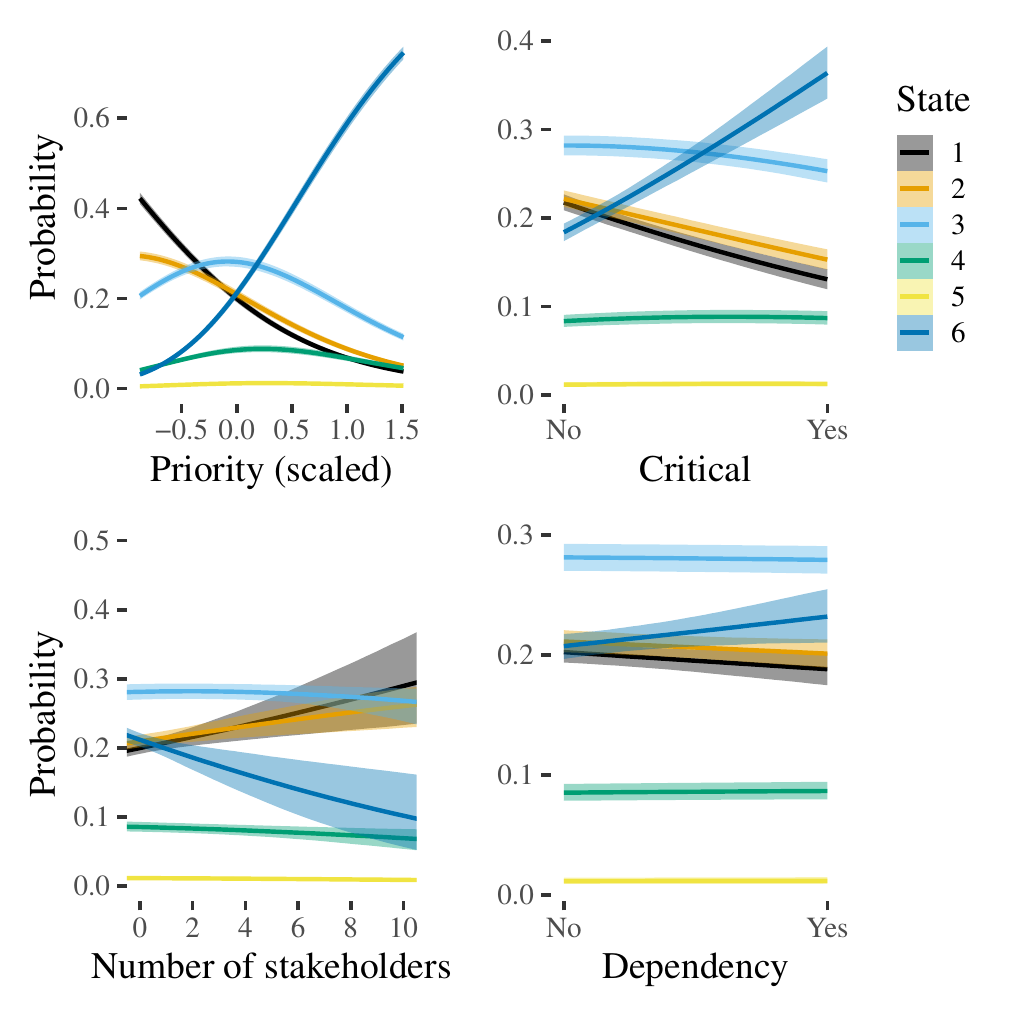}
    \caption{Conditional effects for four of the predictors. On the $y$-axis we have the estimated probability (which can differ between the plots). The top-left plot has a scaled $x$-axis, while for the bottom left plot it is on the outcome scale. The two plots to the right have a dichotomous outcome `No'\slash `Yes' on the $x$-axis. The colored bands show the 95\% credible interval.}
    \label{fig:ce-pop}
\end{figure}

Conditional effects allow us to fix all predictors to their mean, or reference category for factors, except for the one we want to understand better.\footnote{For unordered categorical variables, which we do not use, the first category is the reference category. This can however be changed if needed. All other categories are deviations from this category (as is the case of many model designs, whether frequentist or Bayesian).} If we plot our significant effects for our continuous covariate, one can see how the effect varies depending on \texttt{State} (Fig.~\ref{fig:ce-pop}).

Let us now go through these plots one by one and make notes about the particular characteristics of an effect. The top-left plot in Fig.~\ref{fig:ce-pop} presents the effect \texttt{Priority}. What is evident, compared with the other plots, is that the uncertainty is low since the bands surrounding each line are tightly following the line. If we look at \texttt{State} $6$ (a feature is released), we can see that there is a much higher probability ($y$-axis) for \texttt{State} $6$, as we move to the right (the priority increases). Also, as expected, for \texttt{States} $1$ and $2$ to have a high \texttt{Priority} is uncommon.

Examining the next plot (clockwise), one can see that for \texttt{State} $6$, there is a clear change when moving from `No' to `Yes' (albeit with a slight increase in uncertainty). Next, in \texttt{Dependency}, which is also a dichotomous variable, there is a difference between \texttt{State} $6$ and \texttt{State} $1$ and $2$. In short, if we have a dependency, there is a greater probability that the feature will end up in \texttt{State} $6$, while the opposite holds for \texttt{State} $1$ and $2$. Also worth noting is that \texttt{State} $3$ has the highest probability of having a dependency (and then it does not matter if it moves from `No' to `Yes').

Finally, for \texttt{Stakeholders}, one can see that \texttt{State} $1$ and $2$ implies that the more stakeholders, the higher the probability that it will end up in those states, which might sound counterintuitive; however, we also have an increase in uncertainty. The opposite holds for \texttt{State} $6$, but once again, greater uncertainty when increasing the number of stakeholders. If we examine Fig.~\ref{fig:hist-stake}, we also receive an answer for why the uncertainty increases in this way, i.e., we have less data (evidence) when the number of stakeholders increases.

We will refrain from plotting the last three significant category-specific effects, and simply conclude by saying that all three effects contain categories that affect the outcome positively or negatively. 

To summarize this section, we have seen that analyzing estimates drawn from the posterior probability distribution provides us with indications of significant effects (by looking at the standard deviations and credible intervals). The conditional effects analysis, i.e., fixing all other variables except for one, provided insight into how an effect varies (an effect's size is, by itself, not always exciting, but rather how it varies depending on context).

\section{Discussion}\label{Discussion}
In this section, the results are discussed and related to previously published findings. Section~\ref{Discussion_RQ1} discusses the first research question, while the second research question is discussed in Section~\ref{Discussion_RQ2}. Section~\ref{Discussion_General} discusses general findings and, finally, Section~\ref{Discussion_implprac} discusses implications for practitioners. 

\subsection{RP criteria with actual impact (RQ1)}\label{Discussion_RQ1}
In analyzing RQ1, this section examines which RP criteria the company deemed most important to use in their project (i.e., which ones are actually used), and which ones have an actual impact when deciding which features should be implemented and released. However, we did not investigate if the used RP criteria are the most \textit{appropriate} ones to use. This decision was made by the company and is not part of this study.

Looking into which RP criteria the company deemed most important to use in practice, eight RP criteria were used when prioritizing requirements, namely:

\begin{itemize}
  \item \textbf{Team priority} - the teams subjective\slash expert opinion of the importance of a feature,
  \item \textbf{Critical Feature} - if the feature was critical or not, 
  \item \textbf{Customer value} - how valuable the feature was considered to be for customers,
  \item \textbf{Business value} - the business value of the feature,
  \item \textbf{Stakeholders} - number of key internal stakeholders who considered a feature important,
  \item \textbf{Key customers} - number of key customers who considered a feature important,
  \item \textbf{Dependency} - if a feature has a dependency to other features, and
  \item \textbf{Architects' involvement} - the needed level of involvement from a software architect in order to design\slash implement a feature. 
\end{itemize}

All eight RP criteria used by the case company have already been identified in the literature (e.g., in~\citep{Svensson2011, Riegel2015, Thakurta2017, Hujainah2018, DANEVA2013, ZHANG2014, Eckstein}). Business and customer value are native to agile~\citep{Eckstein}, and used in industry when prioritizing requirements~\citep{Svensson2011,DANEVA2013}. Although expert opinion (peoples' previous experiences, opinions, intuitions, various criteria, arguments, or a combination of one or several of these information sources) is not identified as an RP criterion in the literature, it is often used when prioritizing requirements~\citep{Svensson2011, Walid2016, Olsson2014}. However, there is a difference between how expert opinion is used in the analyzed project and what is reported in the literature. The difference is that the team's expert opinion (called Team priority) is an explicitly specified criterion for RP where the teams decide on a value (between 0 and 1000) that represents their expert opinion, while in the literature expert opinion is not stated as an explicit RP criterion and it is not quantified.

One interesting finding is related to the importance of a requirement. \citet{Hujainah2018}, indicate that importance is the most frequently used RP criterion in the identified RP techniques\slash tools. According to \citeauthor{Hujainah2018}, importance refers to how important a requirement is to the stakeholders. This definition is in line with~\cite{Thakurta2017}, who defines importance as the subjective evaluation of a requirement by stakeholders. However, stakeholders include several different types of stakeholders, e.g., users, customers, the project team, marketing\slash business department, and competitors; thus it is not clear in~\cite{Hujainah2018, Thakurta2017} which perspective is used. On the other hand, \citet{Riegel2015}, report on several different perspectives of importance identified as RP criteria, e.g., project importance with regards to overall project goal, importance to business goals, and importance to customers. In the analyzed project, three different perspectives of importance were used when prioritizing requirements as three separate criteria, namely: (i) from the project's perspective (Critical feature), (ii) from an internal stakeholder perspective (Stakeholders), and (iii) from a customer perspective (Key customers). The analyzed project's different perspectives of importance is in line with the view of \citet{Riegel2015}. In the literature, importance is often used in pair-wise comparisons, to produce an ordered list of requirements based on importance, or from a cost-value perspective. However, in the analyzed project, the importance from stakeholder and customer perspective were simply used by counting how many internal stakeholders considered the feature\slash requirement to be important and counting how many key customers (customer perspective) consider a feature\slash requirement to be important.

One surprising finding, when comparing the used RP criteria in the analyzed project with the literature, is that implementation\slash development effort\slash cost was not used at all in the analyzed project, despite being frequently mentioned in the literature (e.g., in~\cite{Hujainah2018, Thakurta2017}), and being the most frequently mentioned criterion in~\cite{Riegel2015}. Moreover, there are several RP techniques\slash tools in the literature (e.g., in~\citep{Bukhsh2020, Hujainah2018}) that are based on cost\slash effort, and it has been reported to be used in industry when prioritizing requirements, e.g., in~\cite{Svensson2011, DANEVA2013}. Despite that various cost\slash effort estimations are performed at the case company and for the analyzed project, it is not deemed as an important criterion to be used for RP. One possible explanation may be that cost\slash effort estimations were considered by the team when setting their own priority (called Team priority), but not explicitly used when prioritizing requirements. However, it is not possible to confirm or reject this explanation based on the extracted data. We can only conclude, based on the extracted data, that implementation\slash development effort\slash cost was not considered an important RP criterion at the company when determining which requirements should be implemented and released, which is not in line with the literature.

The case company used eight RP criteria in the analyzed project, but just because they are used it does not mean that they have an actual impact when determining which requirements should be implemented and released. Therefore, we analyzed 32,139 decisions for 11,110 features to see which of the eight RP criteria have an actual impact. Based on the results in Sect.~\ref{Res} (see Table~\ref{tab:summary} and Fig.~\ref{fig:mcmc_areas}), seven out of the eight RP criteria used in the analyzed project have an actual impact on RP. The only criterion that did not have an actual impact was Key customers.

When comparing the RP criteria that do have an actual impact on the RP decisions made, there is a difference between the seven RP criteria in how strong of an impact each criterion had, as shown in Table~\ref{tab:summary} (column Estimate) and in the replication package. Two criteria had a strong impact on the RP decisions, namely Team priority and Critical Feature. Team priority had the strongest impact on the RP decisions (with an estimate of $1.22$), while Critical Feature had an estimate of $0.62$. The remaining five RP criteria had a small impact with an estimate between $-0.29$ and $0.19$. The finding that Team priority (i.e., the teams’ expert opinion\slash experiences\slash subjective opinion) had the strongest impact on RP decisions is in line with the literature~\citep{Olsson2014, Walid2016} which suggest that RP decisions are commonly based on previous experiences and opinions.

Although the RP criterion Dependency was significant, meaning it had an actual impact on RP, it had a low impact on deciding which requirements that were implemented and released, which is shown in Table~\ref{tab:summary} and Fig.~\ref{fig:mcmc_areas}. This result is not in line with the literature~\citep{DANEVA2013, Riegel2015, Thakurta2017, Hujainah2018, SHAO2017, ZHANG2014}. This is surprising since requirement dependencies are important when prioritizing requirements and deciding the order in which the requirements can be implemented~\citep{ZHANG2014}. Some requirement needs to be satisfied according to conditions of other requirements, while others may have to be implemented together~\citep{Li2012}. According to \citet{SHAO2017}, requirement prioritization results that do not consider requirements dependency can rarely be used, which is not in line with the results from this study. In addition, requirement dependencies are used as an RP criterion in industry~\citep{DANEVA2013}, is frequently mentioned as an important RP criterion in the literature~\citep{Riegel2015, Thakurta2017}, and used in several RP techniques\slash tools~\citep{Hujainah2018}. However, just because dependency is used as a RP criterion, it may not have a large impact on the RP decisions made, as shown in this study. One possible explanation for the difference between this study and the literature is that we have not asked industry practitioners what they consider (i.e., their subjective opinion) to be important when prioritizing requirements, nor have we used our own opinion or previous studies to decide which criteria have an actual impact on RP. Instead, we investigated the actual outcome of 32,139 RP decisions for one completed project at one software developing company. To the best of our knowledge, no other study has analyzed the actual outcome of RP decisions in industry to identify which RP criteria have an actual impact, and definitely not with such a large sample.

\subsection{Impact of RP criteria depending on the state (RQ2)}\label{Discussion_RQ2}
As shown in Table~\ref{tab:DiscRQ2} (the conclusions in Table~\ref{tab:DiscRQ2} are based on the results in Sect.~\ref{Res} and in the replication package), different RP criteria had different impact on RP depending on the state of the requirement, i.e., depending on how far the requirement has reached in the development process. Meaning, some criteria had a high impact on RP early in the development process, others in the middle, while some had a high impact at the end.

\begin{table*}
\caption{Requirements prioritization criterion's impact depending on a requirements state- A '-' means no impact.}\label{tab:DiscRQ2}
\begin{tabularx}{\textwidth}{lllllllll}
\hline
\textbf{Cutpoint} & \textbf{Team Priority} &  \textbf{Bus. value} & \textbf{Cust. value} & \textbf{Critical} & \textbf{Depend.} & \textbf{Stakeholders} & \textbf{Key cust.} & \textbf{Arch. involv.}\\
\hline
1 & Low & High & - & No & - & More & - & More\\
2 & Medium & - & - & No & - & - & - & More\\
3 & - & High & High & - & - & - & - & -\\
4 & - & Low & High & - & - & - & - & Less\\
5 & High & - & - & Yes & - & Less & - & Less\\
\hline
\end{tabularx}
\end{table*}

When moving from State~1: \textit{Elicited} to State~2: \textit{Prioritized} (cutpoint 1 in Table~\ref{tab:DiscRQ2}), five RP criteria had an impact on the RP decision. The lower team priority, the higher business value (i.e., to be considered valuable for the company), not being considered a critical feature, the more internal stakeholders that consider a feature to be important, and the more architects are involved, the higher probability that a requirement reach State~2. For cutpoint 2 (when a requirement is moving from State~2: \textit{Prioritized} to State~3: \textit{Planned}), the RP criteria stakeholders and architect's involvement have the same impact as in cutpoint 1. In addition, a medium Team priority (i.e., not too low and not too high) had an impact on the RP in cutpoint 2. When a requirement moves from State~3: \textit{Planned} to State~4: \textit{Implemented} (cutpoint 3 in Table \ref{tab:DiscRQ2}), high customer and business value had an impact on the requirement prioritization, i.e., the higher customer and business value a requirement have, the more likely it is that it will reach State~4. High customer value, low business value, and less involvement from the architects are important for a requirement to move from State~4: \textit{Implemented} to State~5: \textit{Tested} (cutpoint 4 in Table \ref{tab:DiscRQ2}). Finally, in cutpoint 5 (moving from State~5: \textit{Tested} to State~6: \textit{Released}), being a critical feature and considered important for the team (i.e., having high Team priority), and having less internal stakeholders interested in the requirement and less involvement from the architects, increases the probability of the requirement to be released.

Looking into the RP criterion Stakeholders, it is only in cutpoint 1 where more internal stakeholders that consider a feature to be important means that a feature is more likely to reach the next state. For all other cutpoints (i.e., RP decisions), Stakeholders had either no impact on the decisions or it is a lower probability for a feature to be included (i.e., prioritized) with an increasing number of internal stakeholders who consider the feature to be important, which is not in line with the literature~\citep{Riegel2015, Thakurta2017, Hujainah2018}.

That the software architects need to be more involved in the beginning of the project (cutpoints 1 and 2) makes sense since it is important to analyze if the included requirements have any negative impact on the current architecture, and\slash or if the technical debt would increase. However, just because a requirement may have a negative impact on the current architecture and\slash or the technical debt, it does not mean it will not be included in the product. It means that it is important to get this information\slash knowledge from the experts (i.e., the software architects) before making decisions about the requirement.

One interesting, and surprising finding is that only internal value (Business value and Stakeholders) and not external value (Customer value and Key customers) had an actual impact on deciding which features reach State~3: \textit{Planned} (up until cutpoint 2). That is, among all features that were prioritized to be included in the product until State~3, only internal value was considered while the customer perspective was ignored. The criterion Customer value only starts having an actual impact when a feature moves from State~3 to State~4 (cutpoint 3), and from State~4 to State~5 (cutpoint 4), while Key customers did not have an impact when prioritizing features. This means that features in the early phases in the development process with high customer value may not be prioritized to be included if the business value is low. This is not in line with~\citep{DANEVA2013} where the focus is on combining value-creation for the vendor (i.e., Business value) with value-creation for the customer (i.e., Customer value). 

The findings in this study show that the team's expert opinion\slash experiences\slash subjective opinion etc.\ only had a positive impact on RP at the very end of the development process (in cutpoint 5). This is not in line with the literature~\citep{Olsson2014, Walid2016}, which suggests that the decisions and selection of what to include are commonly based on previous experiences, opinions, intuitions, arguments, or a combination of one or several of these information sources. Instead, up until cutpoint 5, the decisions (i.e., RP) were based on the internal stakeholders view of the importance of the feature, business value and finally customer value. However, in cutpoint 5, Team priority had a very large effect (probability mass close to 70\%) on which features should be released.

When discussing RQ1, which RP criteria have an actual impact on RP, we saw that Dependency had a significant impact on RP; however, it was weak due to high uncertainty. When analyzing RQ2, if the impact change depending on which state a requirement is in, we see, in particular in cutpoint 2 (i.e., to reach State~3), that Dependency had an impact with a probability mass of close to 30\%. However, not much changes when the dependency moves from `No' to `Yes', as shown in Fig.~\ref{fig:ce-pop}. We see a similar pattern, although with a lower portability mass, for all other states. One possible explanation may be that all features, regardless if any dependencies to other features have been identified when the RP decisions are made, are treated in the RP decision process as if they have dependencies to other features. Meaning, the practitioners did not consider if the value for the Dependency criterion is 'Yes' or 'No', it was viewed as if there are dependencies, and if the dependencies are discovered later in the software development process they will be able to handle it without any delays. This is supported by \citet{Martakis2013} who found that practitioners in agile software development projects indicated that they were able to deal with dependencies without too much effect for the project, and whether the dependencies were discovered early or late did not have an effect or impact on the project. 

\subsection{General discussion of results}\label{Discussion_General}
The results from RQ2 (see Sect.~\ref{Discussion_RQ2}) show that not all RP criteria have an equal impact on which requirements are prioritized to be included, implemented, and eventually released, and that the impact of a criterion changes depending on where in the software development process a requirement is (refers to the six different states, as described in Table~\ref{tab:FeatureVariables}). For example, Business value had an impact on RP in the early phase, Customer value in the later phases, while being a critical requirement (i.e., Critical feature is `YES') only had an impact in the last phase (State~6). These findings are not in line with how RP techniques\slash tools in the literature are developed~\citep{Hujainah2018, Thakurta2017}. Most, if not all, RP techniques\slash tools select, based on expert opinion, which criteria should be used in the developed RP technique\slash tool~\citep{Riegel2015}. Hence, the RP technique\slash tool in the literature cannot be used in a flexible way with different criteria depending on the development phase, and thus may not be so useful in practice. This may be one reason why gut-feeling, subjective opinion, and expert judgement are frequently reported to be used in RP~\citep{Svensson2011, Walid2016, Olsson2014} and it may be explained by the representativeness heuristic, which is a mental shortcut to lessen the cognitive load~\citep{gren2017possible}. Meaning, when the needed information (i.e., RP criteria with an actual impact) is not available\slash cannot be used in the current tools\slash techniques when making decisions, practitioners use similar or previous experience (e.g., their gut-feeling or subjective opinion) instead. The importance of having flexible RP techniques\slash tools is supported by~\citet{Berander2005}.

The findings in this study show the importance for customizing RP criteria, not only to a specific context\slash project~\citep{Riegel2015}, but also to specific development phases. Thus, when developing RP techniques\slash tools, or other decision support systems (e.g., AI-based or data-driven decision support systems), it is important to identify which criteria are important to use (i.e., which ones have an actual impact on RP decisions), and when to use them. Not identifying which criteria that have an impact may lead to other consequences. One consequence may be that unnecessary criteria (information that is not important for the decision) are presented to the decision makers. That is, unnecessary decision criteria are visible for the decision makers, which may lead to poor or wrong decisions. Having an extra irrelevant option set, e.g., RP criteria that have no impact on the decision, visible to the decision maker should not affect the choice, but in some contexts it does~\citep{huber}. The importance of presenting correct information to the decision maker is shown in ~\citep{gren2017possible} where the presence of obsolete requirements negatively affected the cost/effort estimations of the requirements. Thus, it may have a similar affect on unnecessary RP criteria.

\subsection{Implications for practitioners}\label{Discussion_implprac}
In this section, we discuss the findings most important to industry practitioners. There are two aspects that could be of interest to industry practitioners in RP contexts: 1) not all RP criteria have an equal impact on RP decisions made, and that the impact of a criterion changes depending on where in the software development process a requirement is, and 2) to fully understand what RP criteria have an actual impact on RP, a detailed statistical analysis of used RP criteria in previous projects is needed.

First, we found strong evidence that not all RP criteria are equal (in terms of impact on RP decisions), and that the impact of an RP criterion changes depending on how far a requirement has reached in the software development process, as presented in Sect.~\ref{Discussion_RQ1} and Sect.~\ref{Discussion_RQ2}. Based on the results from our statistical analysis, we identified the need to oversee how RP criteria are decided\slash selected (i.e., which one to use) for RP decisions in projects in industry. Since our results show that different RP criteria have different impact depending on when the RP criterion is used (i.e., where in the software development process), it is recommended to consider development phase-specific RP criteria instead of sticking to the same criteria from the beginning to the end. Hence, our recommendation to industry practitioners is to at least consider that a RP criterion may not have an actual impact on RP decisions throughout the entire software development process, and thus select RP criteria to be used, e.g., in the beginning of the project, in the middle, and at the end of the project.

Which RP criteria that are better to use in the beginning, middle, and at the end of the project is not possible for us to recommend based on the analysis in this paper. There are several reasons for this. We have not analyzed or collected the needed data to see if the eight RP criteria used in the analyzed project are the most appropriate ones to use. Different companies and projects may, or perhaps should use different RP criteria than the ones used in the analyzed project. The selection of which RP criteria to use should depend on the specific project and development phase. For example, our results show that only internal value (business value and stakeholders) and not external value (customer value and key customers) have an actual impact on deciding which requirements reach state 3. However, this may not be what is important for a project or a company. Perhaps external value should first have an impact on the RP decisions (e.g., in the beginning) and then internal value should have an impact (e.g., in the middle of the software development process). If this is the case, the project and the company should not include internal value as RP criteria in the beginning of the project. Therefore, we do not recommend specific RP criteria for specific development phases. Instead, we recommend the practitioners to consider that not all RP criteria are equal, and that there may be a need to oversee how they select RP criteria where it is recommended to consider development phase-specific RP criteria instead of sticking to the same criteria from the beginning to the end.

Second, our work highlights the importance of conducting a detailed statistical analysis of RP criteria used in previous projects to identify which RP criteria have an actual impact on RP decisions made. Not identifying which RP criteria that have an actual impact on RP may lead to negative consequences. For example, if an RP criterion with no actual impact is presented to the decisions makers, it may affect the decisions in a negative way (e.g., poor or wrong decisions are made); and time is wasted on collecting\slash recording values to RP criteria with no actual impact on the decisions. Thus, by conducting a statistical analysis of historical RP criteria's actual impact on RP decisions, a knowledge base can be built to identify which ones have a strong impact on RP decisions and which ones are less important. This knowledge base can then be used for future projects within a company when deciding which RP criteria should be used and where in the software development process they have the strongest impact.

\section{Validity threats}\label{ValidityThreats}
\textit{Construct validity}~\citep{ralphT18constr} is concerned with the relation between theories behind the research and the observations. A construct in this case is a latent concept we are trying to measure. Ultimately we want to measure if the concept is real and if, the way we measure it indirectly, is appropriate to better understand the concept.

Concerning this study, the variables (e.g., RP criteria) used in the statistical analysis are all constructs and, hence, try to measure an underlying latent concept; this served our purposes well. By investigating literature and then contrast this with our statistical analysis we uncovered several cases where one could question if appropriate constructs are used in RP. First, the effect between constructs vary, which might not be a problem by itself; however, the fact that they vary over time is a bit more worrying. This could be a sign of inappropriate constructs (Sect.~\ref{Discussion_RQ2}) and we have in this study taken a first systematic step to analyze these constructs using a principled approach to statistical analysis (see, e.g., early work by~\citet{furiaTF2021applying}).

In summary, for our constructs, there might be face validity (does it makes sense?); however, content validity (do the constructs include all dimensions?) is most likely lacking for some constructs. This indicates that predictive validity can be questioned.

\textit{Internal validity} concerns whether causal conclusions of a study are warranted or if overlooked phenomena are involved in the causation. We assume that the eight used RP criteria in the analyzed project are considered to be the most important ones to use when prioritizing requirements. However, other RP criteria than the ones in the database may have been used and, thus, affected the results. Moreover, another factor that may have affected the results is incorrect\slash missing data\slash value for the different RP criteria. There were no NAs in the dataset; however, that does not necessarily mean that there are no NAs. Some of the coding can be a representation of NA, e.g., `No Value'. In this case, we know that `No value' and `None' in the dataset actually are values and not a representation of NAs since we asked the "gate-keeper" from the analyzed project about the correctness of the data.

\textit{External validity} is concerned with the ability to generalize the results, i.e., in this case the applicability of the findings beyond the studied project and company. Analytical generalisation enables drawing conclusions and, under certain conditions, relating them to other cases. This means that the context of the study needs to be compared to the context of interest for the findings to be generalised to. Therefore, we describe the case company and the studied project in as much details as possible considering confidentiality (see Sect.~\ref{subsec:casecompany}). However, the results of which RP criteria have an actual impact on RP decisions, and if their impact changes depending on how far a requirement has reached in the software development process, is specific for the studied project and the case company. 

Even though we analyzed 11,110 features and 32,139 RP decision based on 8 RP criteria, we only analyzed one completed project. Thus, it is possible that the results would have been different if we studied other projects, RP decisions, and RP criteria. Our study was not designed to develop theories that applies to all projects, RP criteria and decisions, but rather to identify trends that could be a first step towards new knowledge and theories. However, it is not possible to generalize the results from this study; although from a transferability perspective, the results may provide an overview that not all RP criteria have an equal impact on RP decisions, and that the impact changes depending on where in the software development process RP decisions are made. Meaning, for projects with a similar context, i.e., projects with thousands of requirements to be prioritized using several RP criteria in many different prioritization points by several practitioners\slash development teams, we would expect to see similar results in terms of: (1) that not all used RP criteria have an actual impact on the decisions throughout the entire software development process, (2) that some RP criteria have a higher impact than others in the beginning while others have higher impact at the end, and (3) that some RP criteria may not have an impact in one or several phases in the development process. However, we do expect to see changes in which specific RP criteria that have an impact on the decisions, both overall and for the specific development phases (beginning, middle, and end). One reason for this is that not all projects in all contexts will use exactly the same eight RP criteria as the analyzed project in this study. Even if some of the eight RP criteria would be used in other projects, they may be used differently.

\section{Conclusion}\label{Conclusion}
In conclusion, we conducted a quantitative study where quantitative data was collected through a case study to analyze which RP criteria have an actual impact on RP decisions, and if the impact of a criterion changes depending on how far a requirement has reached in the development process. To this aim, we extracted 32,139 RP decisions based on eight RP criteria for 11,110 requirements (features) from one completed project at a case company. The extracted data was analyzed by designing, comparing, validating, and diagnosing ordinal Bayesian regression models. We showed how to model ordinal data in a principled way, how to use category-specific effects to get a more nuanced view, and how to report results using conditional effects. The results from this study highlights the following key findings:

\begin{enumerate}
    \item Not all used RP criteria have an actual impact on RP decisions, e.g., Key customers had a very slight positive effect, which was not significant according to the 95\% credible interval.
    \item Not all RP criteria have an equal impact, and this changes depending on how far a requirement has reached in the development process. For example, for RP decisions before iteration\slash sprint planning, having high Business value had an impact on RP decisions, but after iteration\slash sprint planning having high Business value had no impact. Moreover, high Team priority (i.e., the teams' subjective opinion) and being a critical feature (i.e., Critical feature is `YES') only had an impact at the very end of the development process. 
    \item Internal value (Business value and Stakeholders) is more important (i.e., have an actual impact on decisions) than external value (Customer value and Key customers) when prioritizing requirements in the beginning of the project. That is, among all requirements that are prioritized to be included in the project until the iteration\slash sprint planning meeting, only internal value is considered, while the customer perspective is ignored.
    
    \item Although Dependency was found to have a significant impact on RP decisions, in particular in the middle of the development process, not much changes, in terms of actual impact in decisions, when the dependency moves from `NO' to `YES'. Meaning, if a requirement has dependencies to other requirements has no impact on requirement prioritization decisions.
\end{enumerate}

The findings in this paper confirm the need for analyzing and identifying which RP criteria are important in order to develop flexible RP techniques\slash tools~\citep{Berander2005}. That is, the importance for customizing requirement prioritization criteria, not only for specific contexts\slash projects, but also for specific development phases.

Finally, the findings in this study highlights the need for conducting more quantitative studies (preferable in combination with qualitative data) on different projects and contexts, and with different RP criteria in order to get a more complete understanding of which RP criteria have an actual impact in RP decisions, and when in the development process they should be used. Although we only studied RP decisions and criteria, the results that different criteria (data\slash information) have different impact on the decisions depending on where in the development process the decisions are made, may be applicable to other types of decisions within software development. Therefore, it would be interesting to study other types of decisions using other support systems, e.g., AI-based, machine learning, or data-driven decision support systems, to identify which criteria\slash data\slash information have an actual impact on the decisions, and when in the development process.

\section*{Acknowledgments}
The computations were enabled by resources provided by the Swedish National Infrastructure for Computing (SNIC) at Chalmers Centre for Computational Science and Engineering (C3SE), partially funded by the Swedish Research Council through grant agreement no.\ 2018--05973.



  \bibliographystyle{elsarticle-harv} 
  \bibliography{index.bib}





\end{document}